\documentclass[prb,aps,twocolumn, longbibliography,floatfix,10pt]{revtex4-2}
\usepackage[
  colorlinks=true,
  urlcolor=blue,
  linkcolor=blue,
  citecolor=blue,
  bookmarks=false,
  pdftitle={},
  pdfauthor={}
]{hyperref}
\usepackage{import}
\usepackage{physics}
\usepackage{subfiles}
\usepackage{graphicx}
\usepackage{bm}
\usepackage{amsmath}
\usepackage{amssymb}
\usepackage{mathtools}
\usepackage{amsfonts}
\usepackage[all]{xy}
\usepackage{xspace}
\usepackage{accents}
\usepackage{stmaryrd}
\usepackage{tabularx}
\usepackage{extarrows}
\usepackage{float}
\usepackage{multirow}
\usepackage{makecell}
\usepackage[capitalise]{cleveref}
\usepackage[dvipsnames]{xcolor}
\usepackage[normalem]{ulem}
\usepackage{soul}
\usepackage{pdfpages}
\makeatletter
\AtBeginDocument{\let\LS@rot\@undefined}
\makeatother

\begin{document}

\title{Emergent topology in thin films of nodal line semimetals }
\author{Faruk Abdulla}
\affiliation{Physics Department, Technion - Israel Institute of 
Technology, Haifa 32000, Israel}
\affiliation{The Helen Diller Quantum Center, Technion, Haifa 32000, Israel}

\begin{abstract}

We investigate finite-size topological phases in thin films of 
nodal line semimetals (co-dimension 2) in three dimensions. By 
analyzing the hybridization of drumhead surface states, we 
demonstrate that such systems can transition into either a 
lower-dimensional nodal line state (co-dimension 1) or a fully 
gapped trivial phase. Additionally, we explore the hybridization 
of bulk states along the nodal loop due to quantum confinement 
when the system is finite in 
directions parallel to the loop’s plane. This generally results 
in a topologically nontrivial gap. In films finite along 
a single in-plane direction, a partial gap opens, giving rise to 
two-dimensional Weyl cones characterized by a one-dimensional 
$\mathbb{Z}$ invariant. When the system is finite along both in-plane 
directions, a fully gapped phase appears, distinguished by a 
$\mathbb{Z}$ invariant whose value increases with the thickness of the sample. 
We further discuss the bulk-boundary correspondence associated with 
these emergent topological phases.

\end{abstract}

\maketitle

\section{Introduction}

A hallmark of topological insulators and semimetals is the 
presence of robust, boundary-localized states that arise from 
the nontrivial topology of the bulk band structure \cite{Chiu_CTQ_2016, 
Armitage_WDS_2018}. Although 
these states are exponentially localized at the boundaries, they 
exhibit a finite decay length into the bulk. In systems of 
finite size, particularly when the system dimensions become 
comparable to the decay length, boundary states localized on 
opposite surfaces can overlap and hybridize \cite{Zhou_2008}. This hybridization 
can open an energy gap, which is often topologically nontrivial \cite{Potter_Multi_2010, 
Asmar_Topo_2018, He_Topo_2018, Otrokov_Thickness_2019, Chowdhury_Prediction_2019,
Li_Intrinsic_2019, Chao_Magnetized_2020, Liu_Magnteic_2023, Lygo2023, Guo2023, 
Lin2023, Smith2024}, thereby offering a novel route to realize topological phases in 
lower-dimensional systems \cite{Xiao2015, Cook_FST_2023, Cook_TRI_2023, Pal_FST_2025}. 

Recent advances in material fabrication, particularly in the context 
of thin films \cite{Leis_Lifting_2021, Zhang_Crossover_2010, 
Sakamoto_Spectro_2010}, van der Waals heterostructures, and moir\'e systems 
\cite{Geim2013, Hu2020, Chong2018, Kou2014, Hussian2020}, 
have made the exploration of finite-size topological effects 
experimentally relevant and increasingly feasible. While hybridization 
of boundary states in topological insulators typically leads to gapped 
phases in reduced dimensions, finite-size topology (FST) in nodal 
semimetals presents a richer and more intricate landscape.

In nodal semimetals, not only surface states but also bulk nodal 
states can hybridize when the system is made finite along certain 
directions \cite{Xiao2015, Pal_FST_2025}. Recently, the author in 
Ref. \cite{Pal_FST_2025} demonstrated that in a three-dimensional 
Weyl semimetal, hybridization of both the bulk Weyl nodes and the 
Fermi arc surface states can occur in suitably confined geometries. 
The former results in a fully gapped, quasi-two-dimensional Chern 
insulating phase, while the latter leads to a partial gap accompanied 
by the emergence of two-dimensional Weyl nodes.

Motivated by these recent developments in finite-size Weyl semimetals, 
in this work we investigate finite-size topological phases (FSTs) in 
three-dimensional nodal line semimetals (NLSMs). In contrast to Weyl 
semimetals, which feature isolated bulk nodes, NLSMs host one-dimensional 
closed nodal loops in momentum space, along with two-dimensional drumhead 
surface states \cite{Burkov_Topological2011, Schnyder_Ryu_2011, 
Chiu_Classification2014, Fang_Topological2015, Chen2015, Bian_Topological_2016, 
Neupane_Observation_2016, Bzdusek2016, Bzdusek2017, Schoop_DiraCone_2016, Wang2017, Lou2018, 
Belopolski2019, Hosen_2020, Chen2021, Gao2023, Abdulla_SNS_2023, Abdulla_TNL_2023, 
Abdulla_ISP_2025}.

We begin by investigating the hybridization of drumhead surface states 
that originate from opposite surfaces of a thin film. Our analysis reveals 
that this finite-size hybridization can drive the system into one of two 
distinct phases: either a lower-dimensional nodal loop state or a fully 
gapped trivial state. Remarkably, the resulting phase can be predicted 
by a single property of the drumhead surface states in a semi-infinite 
geometry. Specifically, if the wavefunction of the drumhead surface states 
decays in an oscillatory manner-vanishing periodically in space-then 
hybridization gives rise to a lower-dimensional nodal loop state 
which is characterized by a $\mathbb{Z}_2$ invariant. In 
contrast, if such oscillatory decay is absent, the system evolves into 
a fully gapped trivial state.

Furthermore, we explore the hybridization of zero-energy bulk modes when 
the system is made finite along directions parallel to the plane of the 
nodal loop. Throughout this manuscript, the term “finite along a given 
direction” implies that the system is subject to open boundary conditions 
along that direction.  
In films finite along a single in-plane direction, a partial 
gap opens, leading to the formation of two-dimensional Weyl cones, which 
are characterized by a one-dimensional $\mathbb{Z}$ invariant. We demonstrate 
that this $\mathbb{Z}$ invariant captures the existence of edge states 
associated with the emergent two-dimensional Weyl cones when the system is 
further confined (but remains thermodynamically large) along the remaining 
direction.

In contrast, when the system is finite along both in-plane directions 
(wire geometry), all zero-energy bulk modes hybridize, resulting in a 
fully gapped quasi-one-dimensional phase. This gapped phase is also 
characterized by a $\mathbb{Z}$ invariant, whose value can be tuned 
by varying the thickness of the sample.

The remainder of this paper is organized as follows. In Sec. \ref{Sec:FST_Surface}, 
we examine finite-size topological phases arising from the hybridization 
of drumhead surface states, considering systems finite along the direction 
perpendicular to the plane of the bulk nodal loop. In Sec. \ref{Sec:FST_Bulk}, 
we investigate finite-size phases originating from the hybridization of 
zero-energy bulk states, focusing on confinement along in-plane directions.
We also discuss the associated bulk-boundary correspondence. A summary of 
our findings is presented in Sec. \ref{Sec:Dis}.


\section{FST from surface states hybridization}
\label{Sec:FST_Surface}

To demonstrate FST in a three-dimensional NLSM, we consider a 
minimal two bands model defined on a cubic lattice
\begin{align}\label{Eq:HkNLSM}
    H({\bf k}) =  \tilde{v} \left(M_k - \cos{k_za} \right) 
    \tau_x + v_z \sin{k_za} \tau_z,
\end{align}
where $\tilde{v}=v/\sin{k_0a}$, $M_k = (2+\cos{k_0a} -\cos{k_xa} 
- \cos{k_ya})$, and $a$ is the lattice constant. Here $\tau$'s are two by 
two Pauli matrices representing orbital degrees of freedom. Two bands touch on a 
closed loop given by $M_k=1$ which lies in the $k_x$-$k_y$ 
plane at $k_z=0$. The area of the nodal loop increases with 
$k_0$ which lies in the range $0<k_0\le \pi/a$. For small $k_0$, the nodal loop 
approximates to a circle of radius $k_0$. The ratio of the two parameters 
$v_z$ and $v$, which describes group velocities perpendicular to the nodal 
loop along $z$ and $x/y$ direction respectively, plays important role
in the hybridization of the surface states resulting finite size phases. In what 
follows we will assume $sgn(vv_z)=1$ and will set the lattice constant $a=1$.

The NSLM described by Eq. \ref{Eq:HkNLSM} is protected   by chiral symmetry 
${\cal S} H({\bf k}){\cal S}^{-1} =-H({\bf k})$, where ${\cal S}$ is 
realized by $\tau_y$. Because of this chiral symmetry which protects 
the nodal loop against gap opening, the NLSM state can be characterized 
by a one dimensional winding invariant $W$ which is computed on a 
closed loop winding the nodal loop \cite{Zhao_TCS_2013, Matsuura_PBS_2013, 
Chiu_Classification2014, Abdulla_SNS_2023}. The fact that the chirality does 
not act on momenta, chiral symmetry will be preserved for finite size 
system and hence the FST phases which we study below may also be characterized 
by a winding number. Due to bulk-boundary correspondence, 
NLSMs possess drumhead surface states  on open surfaces 
for which nodal loop has a finite projection on the corresponding 
2D surface BZ \cite{Burkov_Topological2011, Schnyder_Ryu_2011, 
Matsuura_PBS_2013, Abdulla_SNS_2023}. For the model defined in 
Eq. \ref{Eq:HkNLSM}, surface states should exist in a  slab geometry which 
is finite along  $z$ direction.

\begin{figure*}
\includegraphics[width=0.9\linewidth]{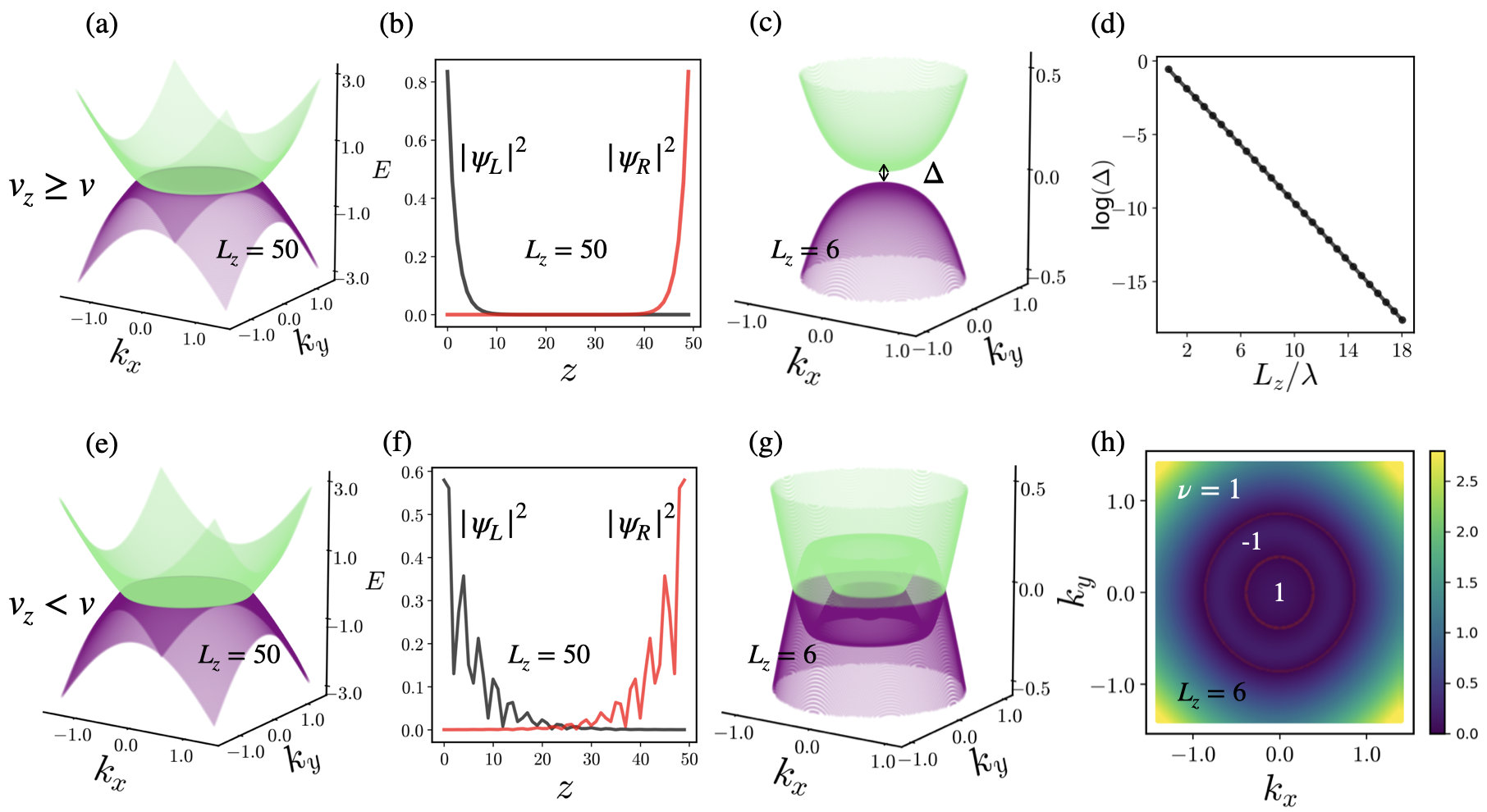}
\caption{Drumhead surface states and finite-size gap in a slab geometry 
finite along the z-direction. The top and bottom rows correspond to $v_z \geq v$
and $v_z < v$, respectively. (a) Low-energy spectrum for $L_z = 50$, 
showing drumhead surface states. (b) Drumhead surface state wavefunctions localized 
at $z = 0$ and $z =L_z=50$  exhibit monotonic decay into the bulk.
(c) Hybridization of these surface states leads to a finite gap $\Delta$. The 
resulting insulating state is topologically trivial. 
(d) Gap decreases exponentially with increasing system size $L_z/\lambda$, where 
$\lambda$ is decay length of the drumhead surface states. Notably, the 
hybridization gap becomes larger as we move away from the center of the 
nodal loop. This is because $\lambda$ diverges near the nodal loop, 
leading to stronger hybridization in its vicinity \cite{Buccheri_Interface_2024}. 
(f) For $v_z < v$, surface state wavefunction exhibits oscillatory decay.
(g) Hybridization in this regime results in a partial gap, with residual 
nodal loops appearing in the quasi-two-dimensional system.
(h) Heatmap of the hybridization gap across momentum space, indicating that 
the spectrum remains gapless along two closed loops (highlighted in red), 
consistent with the analytical prediction (see Eq. \ref{Eq:SNLAna}). The nodal 
loops of the resulting  finite size phase are characterized by a $\mathbb{Z}_2$ 
invariant $\nu$ (see in Sec. \ref{Subsec:TopologyCharact} for details). 
}
\label{Fig:fig1}
\end{figure*}

In the following, we analytically derive the surface states for 
a semi-infinite slab and demonstrate that their decay into the bulk 
is oscillatory when $|v_z| < |v|$, and non-oscillatory when $|v_z| 
\ge |v|$. In the oscillatory regime, hybridization of surface states 
in a finite slab opens only a partial gap, leaving behind zero-energy 
states that form closed loops, resulting in a nodal line phase with 
multiple nodal loops in a quasi-two-dimensional system. Conversely, 
in the non-oscillatory regime, hybridization fully gaps out the 
zero-energy drumhead surface states, yielding a topologically trivial 
insulator in quasi-two dimensions. These results are illustrated 
in Fig. \ref{Fig:fig1}. 

\subsection{Drumhead surface states for semi-infinite slabs}

Let us consider a slab geometry which is periodic along $x$,$y$ directions 
and semi-infinite along $z$ direction. The layers along $z$ direction 
are labeled by an integer $z \in (1, \infty)$. 
This semi-infinite slab of the NLSM in Eq. \ref{Eq:HkNLSM} is described by 
the following Hamiltonian 
\begin{equation}
\begin{aligned}
    H = \sum_{z} \sum_{{\bf k}_{\parallel}} & c^{\dagger}_z({\bf k}_{\parallel}) 
    \tilde{v}M_k \tau_x c_z({\bf k}_{\parallel}) \\
   & - \left( c^{\dagger}_{z+1}({\bf k}_{\parallel}) (\tilde{v} \tau_x + iv_z 
    \tau_z) c_z({\bf k}_{\parallel}) + h.c \right), 
\end{aligned}
\end{equation}
where ${\bf k}_{\parallel}=(k_x, k_y)$. We are looking for eigenstates 
$|\psi_{{\bf k}_{\parallel}}(z) \rangle$ which are decaying into the slab. 
We start with the following ansatz 
\begin{align}
  |\psi_{{\bf k}_{\parallel}}(z) \rangle =  \sum_s u^z \phi_s 
  c^{\dagger}_{zs} |0\rangle,
\end{align}
where $s=(\uparrow,\downarrow)$ represent pseudo-spin degree of freedom 
and $\phi$ is a two-components spinor. The parameter $u$, to be determined, 
is in general complex which must satisfy $|u|<1$ for the wavefunction to 
be normalizable. Plugging the above ansatz in the Schrodinger equation 
$H|\psi_{{\bf k}_{\parallel}} \rangle = E |\psi_{{\bf k}_{\parallel}}\rangle$, 
we find an eigenvalue equation for $\phi$: $P \phi=0$, where the matrix $P$ is
\begin{equation}\label{Eq:Matp}
   P = \begin{pmatrix}
    -2E/\tilde{v} && g_{+} \\
     g_{-} && -2E/{\tilde{v}} 
    \end{pmatrix}, 
\end{equation}
and $g_{\pm} = 2M_k - (u+1/u) \pm  r(u-1/u)$. For nontrivial solution of 
$\phi$, we must have $det(P)=0$ which determines $u$ and is given by 
\begin{subequations}\label{Eq:uval}
\begin{align}
  & u_{\pm} = \frac{\alpha \pm \sqrt{\alpha^2-4}}{2}, \\
  & \alpha = 2 \frac{M_k \pm \sqrt{M_k^2 - (1-r^2)(M_k^2 + r^2 - 
    E^2/\tilde{v}^2)}}{1-r^2},
\end{align}
\end{subequations}
where $r=\frac{v_z}{v}\sin{k_0}$. For a given $u$ the  spinor $\phi$ 
(unnormalized) can be expressed as
\begin{equation} 
\phi = 
\begin{cases}
    \left(1,~ \frac{E/\tilde{v}}{g_{+}}\right)^T & \text{when $g_{+}\neq0$} \\
    \left(\frac{E/\tilde{v}}{g_{-}}, ~ 1\right)^T & \text{when $g_{-}\neq0$}
\end{cases}
\end{equation}
For a given $\alpha$, there are two 
values of $u$ and they satisfy $u_{+} u_{-} = 1$. Clearly there are four 
roots of $u$: Two of them satisfy $|u|< 1$ and the other two $|u|\ge1$. 
For a finite slab $z \in (1, L_z)$, the roots satisfying $|u|<1$ and $|u|>1$ 
would describe surface states which are localized on $z=1$ and $z=L_z$ 
respectively. For semi-infinite slab, the two roots satisfying $|u|<1$ 
allow normalizable solution. Combining these solution, we have a general 
decaying wavefunction $|\Psi_{{\bf k}_{\parallel}}(z)\rangle = c_1u_1^z \phi_1 + 
c_2 u_2^z\phi_2 $ for the semi-infinite slab. The wavefunction 
$|\Psi_{{\bf k}_{\parallel}}(z)\rangle$ must 
satisfy the boundary condition $|\Psi_{{\bf k}_{\parallel}}(0)\rangle={\bf 0}$, which 
(including the normalization condition) determines the unknown constants $c_1$, 
$c_2$ and the energy $E(k_x, k_y)$ of the surface states. Assuming $g_{+}\neq0$, 
we solve the  boundary condition. We find $c_1=-c_2$ and 
$E(k_x, k_y)=0$ for those values of $k_x$-$k_y$ for which we have two roots  
satisfying $|u|<1$. Substituting $E=0$ in the matrix $P$ (see Eq. \ref{Eq:Matp}), 
now the condition $det(P)=0$ reduces to  $g_{-}=0$, which can be easily 
solved for two values of  $u$ 
\begin{align}\label{Eq:Rootu}
    u_{1,2} = \frac{M_k \pm \sqrt{M_k^2 + r^2 -1}}{1+r}. 
\end{align}
For a given $r$, we have 
$|u_{1,2}|<1$ only when (after some straightforward algebra) $M_k<1$ i.e. 
$\cos{k_x} + \cos{k_y} > 1+\cos{k_0}$. Therefore, we find that surface 
states exists for a set of values of $k_x$-$k_y$ which is bounded by 
the projection of the nodal loop on $k_x$-$k_y$ surface BZ. Note that 
all the surface states are at  zero energy. Let us denote the wavefunction 
of the surface states localized on $z=1$  by $|\Psi_{L{\bf k}_{\parallel}}(z)\rangle$. 
We can now explicitly write down the wavefunction for the surface states
\begin{equation}\label{Eq:PsiL}
 |\Psi_{L{\bf k}_{\parallel}}(z)\rangle = C_L \left(u_{1L}^z - u_{2L}^z \right)
 \begin{pmatrix}
     1 \\  0
 \end{pmatrix}
\end{equation}
 for ${\bf k}_{\parallel}=(k_x, k_y)$ satisfying $\cos{k_x} + \cos{k_y} > 
 1+\cos{k_0}$, and $C_L$ is a normalization constant.

\begin{figure}
\includegraphics[width=0.7\linewidth]{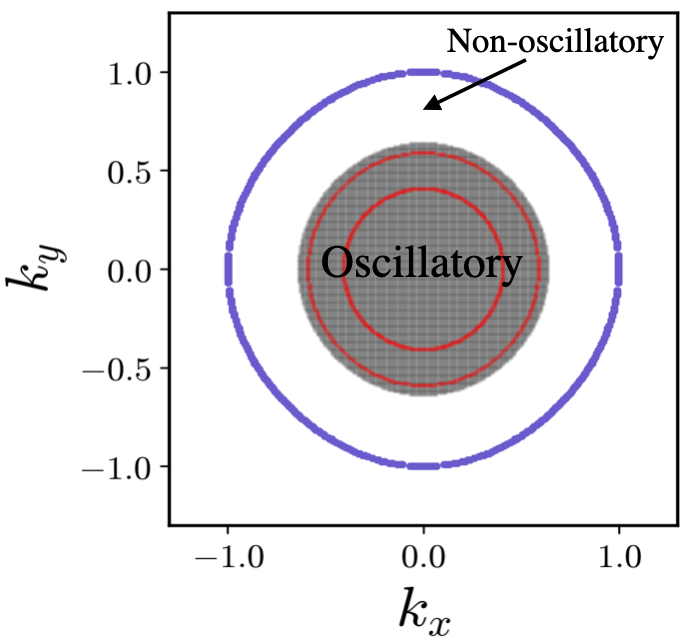}
\caption{Drumhead surface states (for thermodynamically large system) 
exist inside the projection of 
the bulk nodal loop on the $k_x$-$k_y$ surface zone (inside the blue 
loop). For $v_z/v=0.8<1$, only some of the surface modes (shaded region) 
decays in an oscillatory fashion. In a finite size slab $L_z=10$, 
drumhead states which are non oscillatory get fully gapped out. When 
the drumhead surface states are oscillatory, some of them (on red loops) 
does not hybridize and remains gapless.}
\label{Fig:fig2}
\end{figure}

 How do the surface states decay into the bulk? We will shortly see 
 that answering this question is crucial to know  when hybridization 
 of the surface states from opposite surfaces leads a partial or full 
 gap opening in the spectrum. The surface states decay into the bulk 
 in an oscillatory (non oscillatory) fashion when $u_{1L,2L}$ are complex 
 (real). When $u$ is complex, the two roots $u_{1L}$, $u_{2L}$ are complex 
 conjugate of each other (see Eq. \ref{Eq:Rootu}). In this case, the 
 spatial part of the wavefunction can be written as 
 \begin{align} \label{Eq:PsiLSpace}
 u^z_{1L} - u^z_{2L} = \left(\frac{1-r}{1+r}\right)^{z/2} 
 \sin\left(\frac{2\pi z}{\xi_L}\right) 
 \end{align}
where the wavelength $\xi_L$ of periodic oscillation is given by 
$\tan(2\pi/\xi_L) = \sqrt{1-r^2-M^2_k}/M_k$. Therefore, the wavefunction 
vanishes (node) at $2z=n ~\xi_L$, where $n$ is an integer. In a lattice $z$ 
is an integer, so such nodes in the wavefunction appear only when $\xi_L$ 
becomes an integer. 
On the other hand, for non oscillatory case the wavefunction never goes 
to zero because $u_{1,2}$ are reals and they obey $u_1 \neq u_2$. The condition 
for complex $u$ is $M_k^2 + r^2 -1<0$. For a given $r=\frac{v_z}{v} \sin{k_0}$, 
the minimum value of $M_k=2+\cos{k_0} -\cos{k_x} -\cos{k_y}$ is $\cos{k_0}$. 
Plugging the minimum value, we finally find the condition for complex $u$ in 
terms of the parameters of the model i.e. the condition
for oscillatory surface states is $|v_z|<|v|$. Note that for a given 
$v_z, v$ satisfying the above condition, only some of the surface modes 
$(k_x, k_y)$ determined by $M_k^2 + r^2 -1<0$ exhibits oscillatory behavior 
(illustrated in Fig. \ref{Fig:fig2}). 

For large $L_z$, compared to the decay length of the surface states, 
the wavefunction  localized on $z=1$ and $z=L_z$ are 
well approximated by $|\Psi_L\rangle$ and $|\Psi_R\rangle$ respectively. 
$|\Psi_L\rangle$ is given in Eq. \ref{Eq:PsiL}, and 
\begin{equation}\label{Eq:PsiR}
 |\Psi_R(k_x, k_y, z)\rangle = C_R \left(u_{1R}^{z-L_z-1} - u_{2R}^{z-L_z-1} \right)
 \begin{pmatrix}
     0 \\  1
 \end{pmatrix}
\end{equation}
for ($k_x$,$k_y$) satisfying $\cos{k_x} + \cos{k_y} > 1+\cos{k_0}$. Here 
$C_R$ is a normalization constant. Clearly $|\Psi_R\rangle$ satisfies 
only one boundary condition $|\Psi(L_z+1)\rangle={\bf 0}$. For complex 
$u_{1R}, u_{2R}$, the spatial part is now given by 
\begin{align}\label{Eq:PsiRSpace}
 u^z_{1R} - u^z_{2R} = \left(\frac{1+r}{1-r}\right)^{(z-L_z-1)/2} 
 \sin\left(\frac{2\pi (z-L_z-1}{\xi_R}\right), 
\end{align}
where $\xi_R=\xi_L=\xi$. Wavelengths of the $|\Psi_L\rangle$ and 
$|\Psi_R\rangle$ are identical because of the mirror symmetry about 
the $x$-$y$ plane $M_z=\tau_x$: $M_zH(k_x, k_y, k_z)M_z^{-1}= 
H(k_x, k_y, -k_z)$. For a large $L_z=50$, drumhead surface states 
and their wavefunctions in  two different regimes of the ratio 
$v_z/v$ are shown in Fig. \ref{Fig:fig1}.

\vspace{0.5cm}
\subsection{Hybridization of surface states and finite size phases}

For a semi-infinite slab, the surface states of a NLSM lie exactly at 
zero energy. However, in a finite slab of thickness $L_z$, the surface 
states localized on opposite surfaces can hybridize, generally leading 
to the opening of a gap in the spectrum \cite{Buccheri_Interface_2024}. 
While for large but finite $L_z$,
the surface states are well approximated by $|\Psi_L\rangle$ and $|\Psi_R\rangle$, 
these states are not suitable for computing the hybridization energy 
via overlap, as (i) they do not satisfy both boundary conditions, 
$|\Psi(z = 0)\rangle = {\bf 0}$ and $|\Psi(z = L_z + 1)\rangle = {\bf 0}$, 
and (ii) they are orthogonal to each other. A perturbative correction 
to the wavefunctions can be introduced to address these issues and 
estimate the hybridization energy \cite{Okamoto_Onedimensional_2014},
but we do not pursue this approach here. Instead, we construct the 
exact wavefunctions that satisfy the full boundary value problem and 
numerically solve the boundary conditions to obtain the hybridization energy.

For a slab of finite thickness $L_z$, the system occupies the region 
$1\le z\le L_z$, we have to combine all the four solutions of $u$ ($|u|\neq 1$,
see Eq. \ref{Eq:uval}) to construct the wavefunction for the surface 
states,
\begin{align}
   |\Psi_{{\bf k}_{\parallel}}(z)\rangle = c_1u_1^z \phi_1 + c_2 u_2^z\phi_2  + 
   c_3u_3^z \phi_3 + c_4 u_4^z\phi_4. 
\end{align}
The first (last) two terms with $|u_{1,2}|<1$ ($|u_{3,4}|>1$) describes 
exponentially decreasing (increasing) solutions. The wavefunction 
$|\Psi_{\bf k_{\parallel}}(z)\rangle$ contains five unknowns: $(c_1, c_2, c_3, c_4)$ and 
the hybridization energy $E$. These five unknowns are uniquely determined by 
the two boundary conditions $ |\Psi_{{\bf k}_{\parallel}}(0)\rangle =  
{\bf 0}$ and $|\Psi_{{\bf k}_{\parallel}}(L_z+1)\rangle={\bf 0}$, and the 
normalization condition $\langle\Psi_{{\bf k}_{\parallel}}
|\Psi_{{\bf k}_{\parallel}}\rangle=1$. The condition imposed by the 
boundary conditions can be combined into a single matrix equation 
${\bf B} |C\rangle=0$, where 
$|C\rangle=(c_1, c_2, c_3, c_4)^T$ and ${\bf B}$ is a $4 \times 4$ 
matrix.  For nonzero solution of $(c_1, c_2, c_3, c_4)$, the determinant of 
the matrix ${\bf B}$ must vanish which translates into the following condition 
\begin{align} \label{Eq:DetB}
    \begin{vmatrix}
    1 && 1 && 1 && 1 \\
    \frac{1}{g_1} && \frac{1}{g_2} && \frac{1}{g_3} && \frac{1}{g_4} \\
    u_1^{L_z+1} && u_2^{L_z+1} && u_3^{L_z+1} && u_4^{L_z+1} \\
    \frac{u_1^{L_z+1}}{g_1} && \frac{u_2^{L_z+1}}{g_2} && \frac{u_3^{L_z+1}}{g_3} 
    && \frac{u_4^{L_z+1}}{g_4} 
    \end{vmatrix} =0,
\end{align}
where $g_i = 2M_k - (u_i+1/u_i) + r(u_i - 1/u_i)$, $i=1,2,3,4$. Note that the hybridization 
energy $E(k_x,k_y)$ enters in the above condition through $u_i$'s. 

\begin{figure}
\includegraphics[width=0.7\linewidth]{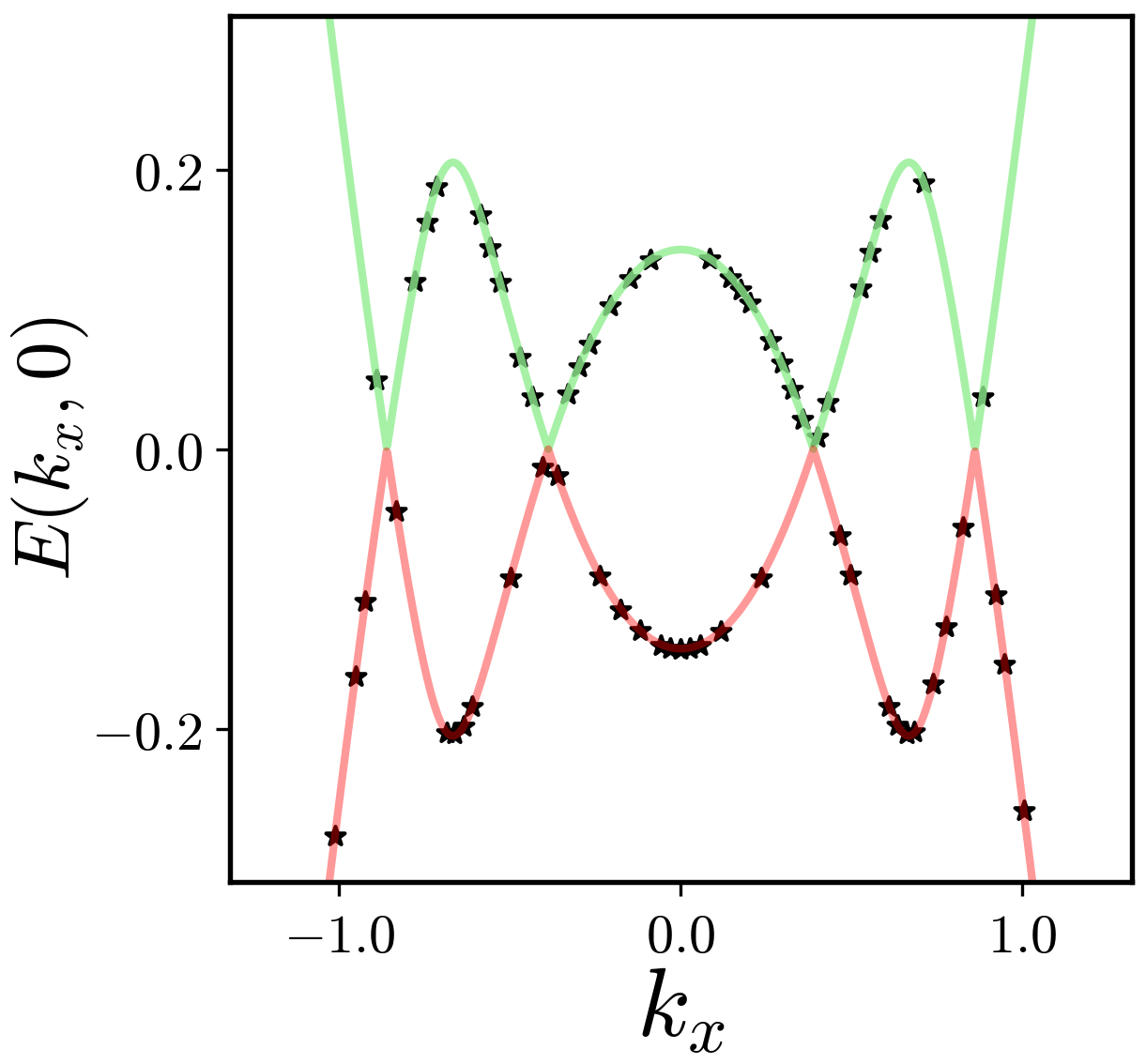}
\caption{Hybridization energy $E(k_x, k_y)$ of drumhead surface states as a 
function of $k_x$ for a fixed $k_y=0$, and $L_z=6$, $k_0=1.0$ and 
$v_z/v=0.2$. Solid lines represent exact diagonalization result and the 
energy computed from Eq. \ref{Eq:DetB} are shown by asterisk.  }
\label{Fig:fig3}
\end{figure}

Alternatively, one can compute the hybridization energy through exact 
diagonalization which involves diagonalization of $2L_z \times 2L_z$ matrix. 
Although we managed to reduce the computational cost by bringing down the problem to 
a $4\times 4$ matrix, the condition Eq. \ref{Eq:DetB} is still not tractable 
analytically  to solve for $E(k_x,k_y)$. We numerically solve the Eq.  \ref{Eq:DetB} 
for  $E(k_x,k_y)$ which agrees with the hybridization energy obtained from 
exact diagonalization as shown in Fig. \ref{Fig:fig3}. We find the following results:  
For a given $(k_x,k_y)$, 
the hybridization gap depends on whether the corresponding surface state 
(in semi-infinite geometry) decays in oscillatory or non oscillatory fashion. 
For the non oscillatory case ($v_z \ge v$), all the surface states hybridize 
and a finite gap opens in the spectrum, thus a trivial insulating state emerges.
On the other hand, for the oscillatory case ($v_z<v$) some of the surface states 
does not hybridize and they remain at zero energy, thus a new semimetallic 
state emerges. The resulting quasi-two-dimensional semimetal hosts multiple 
concentric nodal loops. Topological characterization of this phase will be
discussed in \cref{Subsec:TopologyCharact}.

The fact that the surface state, which decay in an oscillatory fashion, 
sometimes does not hybridize requires an explanation. Although the 
surface states $|\Psi_{L{\bf k}_{\parallel}}(z)\rangle$ and 
$|\Psi_{R{\bf k}_{\parallel}}(z)\rangle$ decay 
exponentially, they retain a finite amplitude (exponential tail) on 
the opposite edges. Consequently, 
for a finite size system, surface states are in general expected to 
hybridize to open a finite gap in the spectrum. However, a special 
situation arises when the surface state contains nodes, meaning the wavefunctions 
$|\Psi_{L{\bf k}_{\parallel}}(z)\rangle$ and 
$|\Psi_{R{\bf k}_{\parallel}}(z)\rangle$ periodically vanish as a function 
of $z$. Reference \cite{Okamoto_Onedimensional_2014} shows 
that there is a simple relationship between the hybridization gap and 
the amplitude of  $\Psi_{L/R}(z)$ at the boundaries $z=0$ and $z=L_z+1$. Note 
that in this context, the terms ``hybridization gap” and ``hybridization 
energy” are used interchangeably. 
They found the hybridization gap $\sim |\Psi_{L/R}(z)|_{z=L_z+1, 0}$. 
It is worth emphasizing that $\Psi_{L/R}(z)$ describes surface 
states in the semi-infinite geometry. Therefore from Eqs. \ref{Eq:PsiLSpace}
and \ref{Eq:PsiRSpace}, we find that 
the hybridization gap vanishes when the wavelength $\xi$ of $\Psi_{L/R}(z)$
satisfies  a condition $n\xi=2(L_z+1)$ which can be rewritten as 
(after plugging the expression of $\xi$)
\begin{align}\label{Eq:SNLAna}
M_k = \sqrt{1-r^2} \cos{\left(\frac{n\pi}{L_z+1}\right)}, ~~ n=1,2,...,L_z. 
\end{align}
The above equation determines the surface modes $(k_x, k_y)$ which 
remain  gapless. Note that for a given $r$ and $L_z$, the above
equation admits solution only for some values of $n$. Clearly for a given
$n$, the modes form a closed loop in the $k_x$-$k_y$ BZ. As an example, 
for $v_z/v=0.2$, $k_0=1.0$ and $L_z=6$,  the Eq. \ref{Eq:SNLAna} has 
solution only for two values of $n=1$ and 2. So there are only two 
concentric closed loops on which the spectrum remain gapless, which 
agrees with both the exact diagonalization results and  the 
solutions of Equation \ref{Eq:DetB}. This is illustrated in 
Figures \ref{Fig:fig1}(g) and \ref{Fig:fig3}.

\subsection{Topological characterization of the FST phase}
\label{Subsec:TopologyCharact}

A nodal loop state in two dimensions  differs fundamentally from 
one in three dimensions  due to their distinct codimensions, defined 
as $\text{codimension} = D - d_N$, where D is the spatial dimension of 
the system and $d_N$ is the dimension of the nodal manifold. For a nodal loop, 
$d_N =1$; thus, its codimension is one in 2D and two in 3D.

A 3D nodal loop state (codimension two) is topologically characterized by 
a one-dimensional invariant, computed over a loop that links the nodal 
line \cite{Zhao_TCS_2013, Matsuura_PBS_2013, Chiu_Classification2014, 
Abdulla_SNS_2023}. In contrast, a 2D nodal loop state (codimension one) is 
characterized by a zero-dimensional invariant, evaluated on a pair of points 
forming a zero-dimensional enclosing manifold—one point $p_1$ inside and 
another $p_2$ outside the nodal loop. This invariant takes values in 
$\mathbb{Z}_2$ \cite{Bzdusek2017, Wu_2018}, and the presence of chiral 
symmetry allows it to be expressed as 
\begin{align}
    \nu = \prod_{i=1,2} sgn(\det Q(p_i)), 
\end{align}
where $Q$ is the upper off-diagonal block of the Hamiltonian of the finite size 
system, in the basis where chirality operator $\cal{S}$ is diagonal. 
The momentum-dependent invariant $\nu$ distinguishes between different gapped 
regions in momentum space. When the two reference points $p_1$ and $p_2$ are 
adiabatically connected-i.e., not separated by any gapless region- we find $\nu = 1$, 
indicating a topologically trivial phase. In contrast, when $p_1$ and $p_2$ 
are separated by a gapless region, $\nu = -1$, corresponding to a topologically 
nontrivial phase.

The quasi-two-dimensional semimetal phase identified in this work features 
multiple concentric nodal loops in the $k_x$-$k_y$ plane of the Brillouin zone. 
For example, as mentioned before, at $v_z/v = 0.2$ and $L_z = 6$, two concentric 
nodal loops are present. This configuration partitions the Brillouin zone into 
three distinct gapped regions: the outermost region (outside both loops) and the 
innermost region (enclosed by the inner loop) both have $\nu = 1$, indicating 
trivial topology, while the intermediate region (between the loops) is characterized 
by $\nu = -1$, indicating a nontrivial topology. These regions are illustrated in 
Figure \ref{Fig:fig1}(h).


\section{FST from bulk states hybridization}
\label{Sec:FST_Bulk}

Finite size effects in topological insulators and semimetals 
introduce coupling between the degenerate and gapless surface 
states localized on opposite surfaces, lifting the degeneracy 
and resulting in a gapped state \cite{Xiao2015, Cook_FST_2023,
Cook_TRI_2023, Pal_FST_2025}. Similarly, in 
topological semimetals, the bulk states at the nodal points 
(band touching points) are degenerate. Finite-size effects 
can couple these bulk nodes, lifting their degeneracy and potentially 
giving rise to a new gapped state \cite{Xiao2015, Pal_FST_2025}. 
For instance, coupling of Weyl nodes due to finite size effects in WSM 
leads to a fully gaped quantum anomalous Hall state in quasi 
two-dimension \cite{Pal_FST_2025}.

In NLSMs that are finite along a direction 
perpendicular to the plane of the nodal loop, only the surface states 
can hybridize as we have seen in the previous section. However, if 
the NLSM is finite along in-plane directions, the degenerate bulk 
nodes-located at different momenta along the nodal loop can couple 
due to quantum confinement, leading to potential gap opening.

In the following, we investigate finite-size effects due to hybridization of 
bulk states in the NLSM described by Eq. \ref{Eq:HkNLSM}. We first consider 
the system finite along a single in-plane direction, say the 
y-direction. In this case, the bulk nodal states at different $k_y$ 
momentum can hybridize. This hybridization results in only a partial 
gap opening, giving rise to isolated Weyl nodes in the spectrum of the 
resulting quasi-two-dimensional system. In contrast, when the system is 
finite along both in-plane directions (i.e., $x$ and $y$), all bulk nodal 
states hybridize, leading to a fully gapped phase. The following sections 
are dedicated to analyzing the finite-size phases that emerge in each 
of these scenarios.

\begin{figure}
\includegraphics[width=1\linewidth]{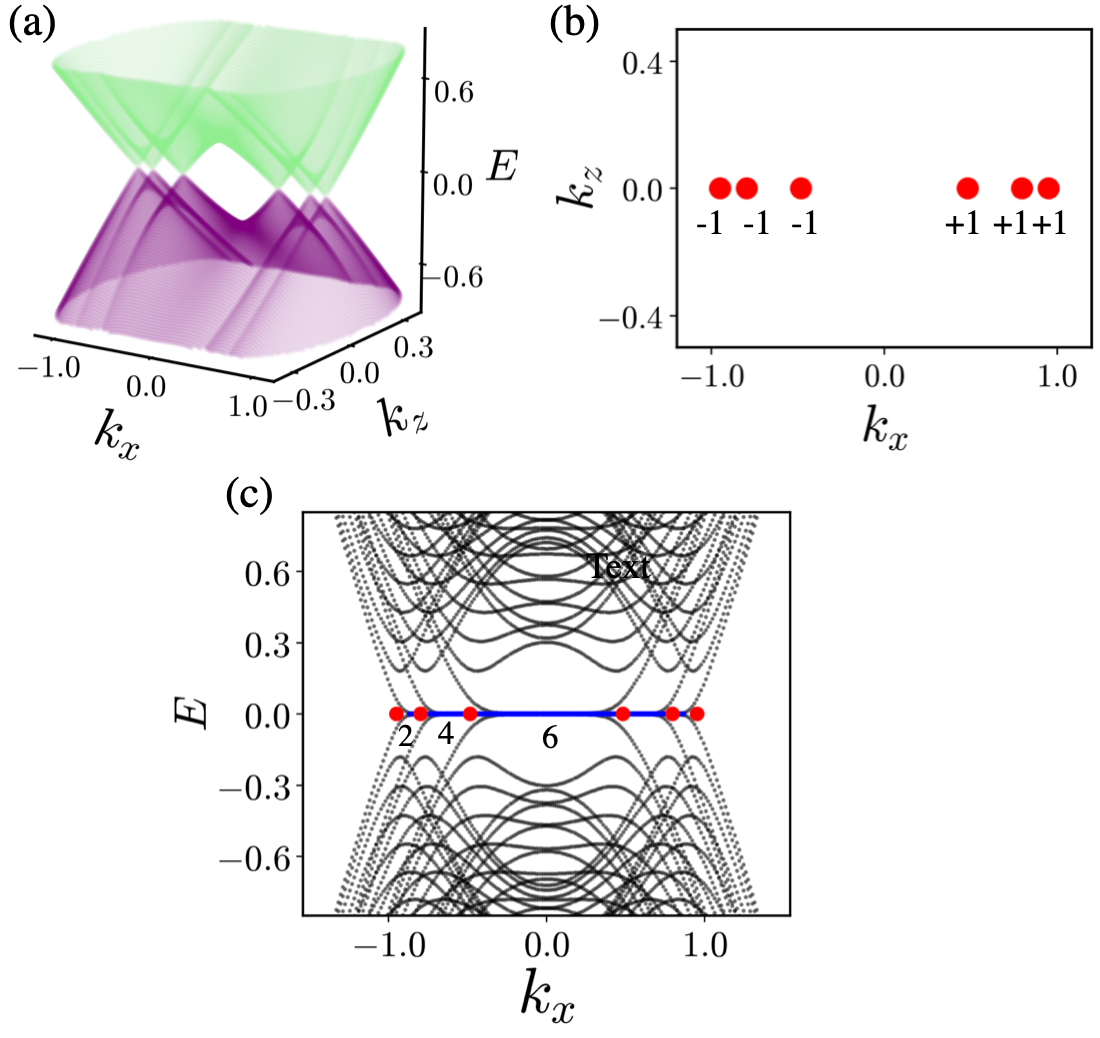}
\caption{(a) For a system finite along the y-direction, hybridization 
of bulk nodal loop states leads to partial gap opening and the 
emergence of Weyl cones in the low-energy spectrum. Parameters are 
$k_0=1.0$, $L_y=10$ and $v_z/v=1.0$. (b) Momentum-space locations 
of the emergent Weyl nodes along with their associated winding numbers.
(c) Edge states (highlighted in blue) corresponding to the Weyl nodes 
(marked in red), exponentially localized near the $z = 1$ and $z = L_z = 50$
boundaries ($L_z \gg L_y$). In the thermodynamic limit along the $x$ and $z$
directions, the degeneracy  of these  edge states (indicated just below edge states)
across different $k_x$ values is consistent with the $\mathbb{Z}$ 
classification of the Weyl nodes. }
\label{Fig:FSTy}
\end{figure}

\subsection{Finite along single in-plane direction -slab geometry}

When the system is taken finite along y direction $y \in (1, L_y)$, 
the bulk nodes at different $k_y$ momentum  hybridize which can leads 
to gap opening. Since there are no additional surface states localised on $y=1$ 
and $y=L_y$ open surfaces, the energy spectrum of the slab can be obtained 
by particle in a box method (PiBM) \cite{Kentaro_Quantum_2021}. The energy 
spectrum in the 
slab geometry can be obtained from the energy spectrum of the system 
in the thermodynamic limit by replacing the momentum $k_y \to n\pi/L_B$, 
where $L_B$ is the length of the box. Note that the boundary condition 
requires the wavefunction to vanish at $y=0$ and $y=L_y+1$. Therefore, 
the length of the box is $L_B=(L_y+1)a$, where $a=1$ is the lattice constant. 
Now we can write down the energy spectrum of the system in slab geometry 
\begin{align}
    E^{\pm}_n(k_x, k_z) = \pm \tilde{v}\sqrt{(M_n - \cos{k_z})^2 + r^2 \sin^2{k_z}},
\end{align}
where $M_n=2+\cos{k_0} - \cos(\frac{n\pi}{L_y+1}) - \cos{k_x}$, and 
$n=1, 2, ..., L_y$. Clearly, valence and conduction bands touch at isolated 
points (nodes) at the zero energy if the first term inside the square root 
vanishes for $k_z=0$ or $\pi$. 
We find that the first term can vanish only for $k_z=0$. Therefore, 
the condition for having nodes in the spectrum reduces to 
\begin{align}
    \cos{k_x} = 1 + \cos{k_0}- \cos(\frac{n\pi}{L_y+1}). 
\end{align}
For a given $k_0$ and $n$, the above equation has solution 
for a pair of nodes only when the right hand side lies in the 
range (-1, 1). Note that the above equation admits solution 
only for some values of $n$. For instance, when $k_0=1.0$ and
$L_y=10$, we have solution only for three values of $n=1,2$ 
and 3. Therefore, there are only six nodes  which are 
separated along  $k_x$ axis, as shown in Fig. \ref{Fig:FSTy}(a).  
Similar to Weyl nodes in three dimensions, these 
two-dimensional nodes are twofold degenerate, possess a topological 
charge, and give rise to Fermi-arc edge states in the host material. 
Consequently, they are sometimes referred to in the literature as Weyl 
nodes \cite{Isobe_Nagaosa_2016, Meng_Zhang_2021, Shi_Liu_2021, Li_Wang_2021, 
Wei_Tao_2022, Jia_Liu_2020, You_Su_2019, Lopes_2024, Stemmer_2023, 
Abdulla_Protected_2024}. A two-dimensional Dirac node, however, is distinguished 
from a Weyl node by the nature of the band degeneracy at the touching 
point.

Recall that the NLSM model described by Eq. \ref{Eq:HkNLSM} 
possesses chiral symmetry, which remains intact even in the 
finite-size system. As a result, the Weyl nodes that emerge in 
the quasi-two-dimensional geometry can be characterized by an 
integer-valued $\mathbb{Z}$ winding number $W$ \cite{Abdulla_Protected_2024}. 
For instance, when $k_0 = 1.0$ and $L_y = 10$, we find six Weyl 
nodes in the spectrum. Our calculations show that each of these 
nodes carries a winding number with magnitude $|W| = 1$ and the their 
sum $\sum_i W_i =0$ due to the Nielson-Ninomiya theorem \cite{Nielsen_Ninomiya_1983}. 
Since the Hamiltonian belongs to symmetry class BDI, Weyl nodes related by 
time-reversal operation carry opposite winding numbers \cite{Abdulla_Protected_2024}.

According to the bulk-boundary correspondence, we expect Fermi arc 
edge states to appear when the quasi two-dimensional 
system is taken finite  along the z-direction 
($L_z \gg L_y$) \cite{Abdulla_Protected_2024}. We have 
numerically computed these edge states associated with the 2D Weyl nodes, 
and the results presented in Fig. \ref{Fig:FSTy}(c) are consistent with 
the $\mathbb{Z}$ classification of the Weyl nodes. 

\begin{figure}
\includegraphics[width=0.9\linewidth]{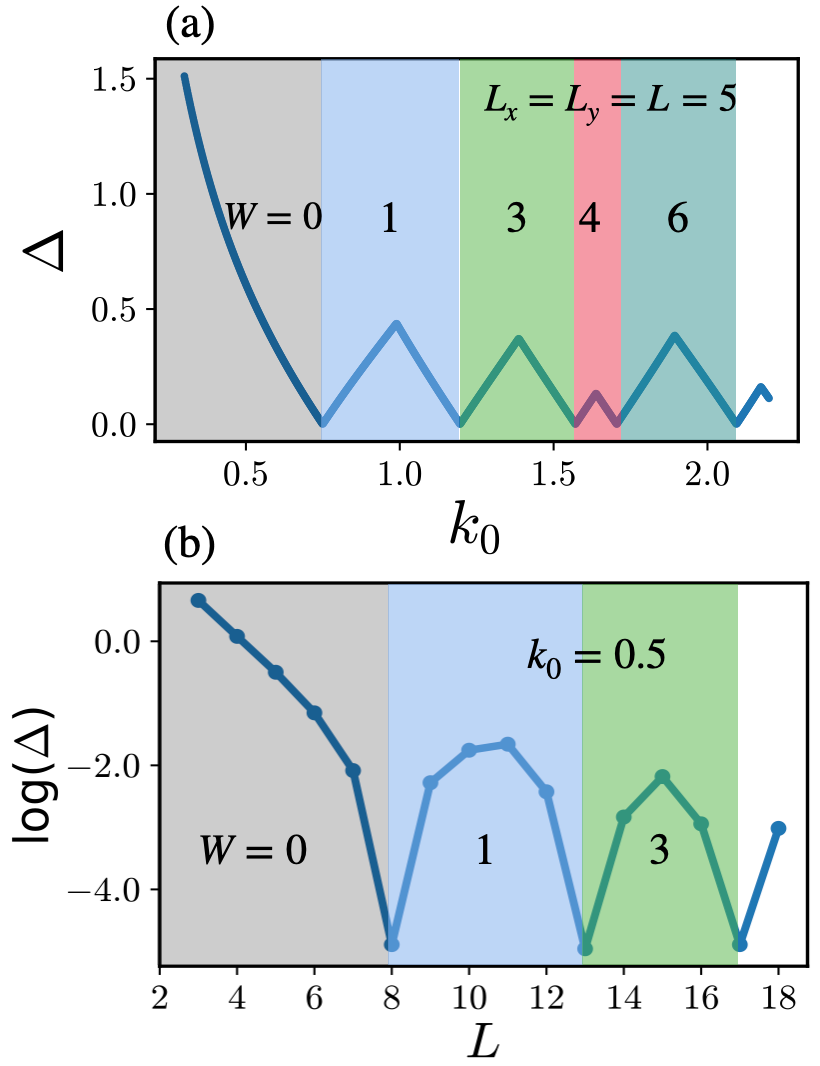}
\caption{When the system is finite along both in-plane directions 
($x$ and $y$), a fully gapped phase emerges. (a) Gap $\Delta$ as a function of 
$k_0$ for a fixed system size $L_x=L_y=L=5$. (b) Log of the energy 
gap as a function of system size $L$ for a fixed $k_0=0.5$. The resulting 
insulating phase is characterized by a winding number  $W$. Every gap 
closing-reopening transition is  accompanied by a change in the winding 
number, reflecting the topological nature of the quasi-one-dimensional state. }
\label{Fig:FSTxy}
\end{figure}

We notice that the gap opening behavior from the hybridization 
of bulk nodal states is independent of the values of $v_z/v$. 
This is in contrast with the case of surface states where 
hybridization and gap opening behavior crucially depends on 
the ratio $v_z/v$. This is due to the fact that the nodal 
loop lies in $k_x$-$k_y$ plane at $k_z=0$, so the group velocity 
$v_z$ of nodal fermion which is perpendicular to the plane 
of nodal loop does not play any role in the hybridization of 
bulk nodal states.

\subsection{Finite along both the in-plane directions -wire geometry}

In Weyl and Dirac semimetals, bulk nodes become coupled when the system 
is made finite along the direction separating the nodes, leading to the 
opening of a topologically nontrivial finite-size gap \cite{Xiao2015, Pal_FST_2025}. 
Similarly, for a NLSM confined along both in-plane directions (wire geometry), 
hybridization of the bulk nodal states is generally expected to produce a 
fully gapped phase with nontrivial topology.

Employing PiBM (discussed in previous section), energy spectrum of 
the finite size system in wire geometry is given by 
\begin{align}
    E_{m,n}^{\pm}(k_z) = \pm \tilde{v} \sqrt{(M_{mn}-\cos{k_z})^2 + r^2 \sin^2{k_z}}, 
\end{align}
where $M_{mn}=2+\cos{k_0} - \cos(\frac{m\pi}{L_x+1}) - 
\cos(\frac{n\pi}{L_y+1})$, and $m=1, 2, ..., L_x$, $n=1, 2, ..., L_y$. 
For a given $k_0$, $L_x$, $L_y$, the spectrum is generically gapped 
and the system is a topologically nontrivial insulator in quasi one-dimension. 
As we have mentioned before, the Hamiltonian of the finite size system 
respects chiral symmetry. Therefore the emerging insulating state 
can be characterized by an one-dimensional $\mathbb{Z}$ invariant which is 
a winding number \cite{Chiu_CTQ_2016}
\begin{align}
    W = \frac{1}{2\pi i} \int_0^{2\pi} dk_z \mathrm{Tr}(Q^{-1} \partial_{k_z}Q), 
\end{align}
where $Q$ is the upper off-diagonal block of the  Hamiltonian of 
the finite size system, in the basis where chirality operator
${\cal S}$ is  diagonal. 

Note that if $k_0$, $L_x$, and $L_y$ satisfy the following condition 
\begin{align}
    1 + \cos{k_0} - \cos\left(\frac{m\pi}{L_x+1}\right) - 
    \cos\left(\frac{n\pi}{L_y+1}\right) = 0, 
\end{align}
the system becomes gapless. For a fixed $(L_x, L_y)$, the finite 
size system goes through a sequence of gap closing and reopening 
transitions as $k_0$ is varied. This is shown in Fig. \ref{Fig:FSTxy}. 
Every gap closing-reopening transitions is accompanied by a change 
in  the winding number of the quasi one dimensional system. We recall 
that the parameter $k_0$ determines the area of the nodal loop in the 
thermodynamic system and the area increases with increasing $k_0$. We
find that the winding number of the quasi one-dimensional system 
increases  with increasing the area of the 
nodal loop in the thermodynamic system (see Fig. \ref{Fig:FSTxy}). 
Similarly for a fixed $k_0$, the finite size system goes through a 
sequence of gap closing and reopening  transitions as ($L_x, L_y)$ i.e. 
thickness of the wire is varied. Therefore, the winding number of the 
quasi one-dimensional system can be tuned by changing the system size. 
Similar behavior has also been reported for Weyl and Dirac semimetal in the 
literature \cite{Xiao2015, Pal_FST_2025}.

By bulk-boundary correspondence, the system hosts edge states which are 
exponentially localized at $z=1$ and $L_z$. These edge states are $W$ 
fold degenerate when the system is thermodynamically large along 
$z$ direction but finite along $x, y$ directions.

\section{Discussion and Conclusion}
\label{Sec:Dis}

In this work, we have demonstrated that finite-size effects in NLSMs
give rise to a rich array of emergent topological
phases. By systematically analyzing both surface and bulk state 
hybridizations under various confinement geometries, we have uncovered 
two distinct mechanisms by which topology is modified in reduced dimensions.

Our study of drumhead surface states revealed that their hybridization 
can result in either a trivial gapped phase or a quasi-two-dimensional 
nodal loop semimetal, depending critically on whether the surface 
wavefunctions decay in an oscillatory or monotonic fashion. This behavior 
is governed by the ratio $v_z/v$, which determines the spatial structure 
of the surface states. We showed that when the surface states decay 
oscillatory, their hybridization can be suppressed at certain thicknesses, 
resulting in multiple, gapless nodal loops whose positions in the Brillouin 
zone are analytically predictable.

In addition to surface state hybridization, we examined  how 
in-plane confinement influences the bulk nodal loop. For slabs 
finite in one in-plane direction, we found the emergence of two-dimensional 
Weyl cones protected by an integer-valued $\mathbb{Z}$ invariant, confirmed 
by the appearance of edge states under further confinement. When the system 
is finite along both in-plane directions, the nodal loop becomes fully 
gapped and the resulting quasi-one-dimensional phase is characterized by 
a  winding number. Importantly, the winding number increases with 
both the thickness of the film and the size of the original nodal loop, 
offering a concrete mechanism to engineer topological invariants via 
geometric control.

Our results expand the conceptual landscape of finite-size 
topology \cite{Xiao2015, Cook_FST_2023, Cook_TRI_2023, Pal_FST_2025} by 
showing how nodal semimetals with lower co-dimension nodes (such as nodal loops) 
respond to geometric confinement in qualitatively distinct ways compared to 
Weyl semimetals. These findings highlight the potential of engineering
lower-dimensional topological states by tuning film thickness and 
symmetry-preserving confinement directions.

Future directions include the study of symmetry-breaking perturbations 
and experimental validation in thin films of NLSM materials such as 
PbTaSe$_2$ \cite{Bian_Topological_2016} and the ZrSiS 
family \cite{Neupane_Observation_2016, Schoop_DiraCone_2016, Lukas_Modular_2020}. 
Our work provides a foundation for exploring finite-size-induced 
topology in systems with higher-dimensional nodal structures.



\bibliography{main}

\begin{thebibliography}{64}%
\makeatletter
\providecommand \@ifxundefined [1]{%
 \@ifx{#1\undefined}
}%
\providecommand \@ifnum [1]{%
 \ifnum #1\expandafter \@firstoftwo
 \else \expandafter \@secondoftwo
 \fi
}%
\providecommand \@ifx [1]{%
 \ifx #1\expandafter \@firstoftwo
 \else \expandafter \@secondoftwo
 \fi
}%
\providecommand \natexlab [1]{#1}%
\providecommand \enquote  [1]{``#1''}%
\providecommand \bibnamefont  [1]{#1}%
\providecommand \bibfnamefont [1]{#1}%
\providecommand \citenamefont [1]{#1}%
\providecommand \href@noop [0]{\@secondoftwo}%
\providecommand \href [0]{\begingroup \@sanitize@url \@href}%
\providecommand \@href[1]{\@@startlink{#1}\@@href}%
\providecommand \@@href[1]{\endgroup#1\@@endlink}%
\providecommand \@sanitize@url [0]{\catcode `\\12\catcode `\$12\catcode
  `\&12\catcode `\#12\catcode `\^12\catcode `\_12\catcode `\%12\relax}%
\providecommand \@@startlink[1]{}%
\providecommand \@@endlink[0]{}%
\providecommand \url  [0]{\begingroup\@sanitize@url \@url }%
\providecommand \@url [1]{\endgroup\@href {#1}{\urlprefix }}%
\providecommand \urlprefix  [0]{URL }%
\providecommand \Eprint [0]{\href }%
\providecommand \doibase [0]{https://doi.org/}%
\providecommand \selectlanguage [0]{\@gobble}%
\providecommand \bibinfo  [0]{\@secondoftwo}%
\providecommand \bibfield  [0]{\@secondoftwo}%
\providecommand \translation [1]{[#1]}%
\providecommand \BibitemOpen [0]{}%
\providecommand \bibitemStop [0]{}%
\providecommand \bibitemNoStop [0]{.\EOS\space}%
\providecommand \EOS [0]{\spacefactor3000\relax}%
\providecommand \BibitemShut  [1]{\csname bibitem#1\endcsname}%
\let\auto@bib@innerbib\@empty
\bibitem [{\citenamefont {Chiu}\ \emph {et~al.}(2016)\citenamefont {Chiu},
  \citenamefont {Teo}, \citenamefont {Schnyder},\ and\ \citenamefont
  {Ryu}}]{Chiu_CTQ_2016}%
  \BibitemOpen
  \bibfield  {author} {\bibinfo {author} {\bibfnamefont {C.-K.}\ \bibnamefont
  {Chiu}}, \bibinfo {author} {\bibfnamefont {J.~C.~Y.}\ \bibnamefont {Teo}},
  \bibinfo {author} {\bibfnamefont {A.~P.}\ \bibnamefont {Schnyder}},\ and\
  \bibinfo {author} {\bibfnamefont {S.}~\bibnamefont {Ryu}},\ }\bibfield
  {title} {\bibinfo {title} {Classification of topological quantum matter with
  symmetries},\ }\href {https://doi.org/10.1103/RevModPhys.88.035005}
  {\bibfield  {journal} {\bibinfo  {journal} {Rev. Mod. Phys.}\ }\textbf
  {\bibinfo {volume} {88}},\ \bibinfo {pages} {035005} (\bibinfo {year}
  {2016})}\BibitemShut {NoStop}%
\bibitem [{\citenamefont {Armitage}\ \emph {et~al.}(2018)\citenamefont
  {Armitage}, \citenamefont {Mele},\ and\ \citenamefont
  {Vishwanath}}]{Armitage_WDS_2018}%
  \BibitemOpen
  \bibfield  {author} {\bibinfo {author} {\bibfnamefont {N.~P.}\ \bibnamefont
  {Armitage}}, \bibinfo {author} {\bibfnamefont {E.~J.}\ \bibnamefont {Mele}},\
  and\ \bibinfo {author} {\bibfnamefont {A.}~\bibnamefont {Vishwanath}},\
  }\bibfield  {title} {\bibinfo {title} {Weyl and dirac semimetals in
  three-dimensional solids},\ }\href
  {https://doi.org/10.1103/RevModPhys.90.015001} {\bibfield  {journal}
  {\bibinfo  {journal} {Rev. Mod. Phys.}\ }\textbf {\bibinfo {volume} {90}},\
  \bibinfo {pages} {015001} (\bibinfo {year} {2018})}\BibitemShut {NoStop}%
\bibitem [{\citenamefont {Zhou}\ \emph {et~al.}(2008)\citenamefont {Zhou},
  \citenamefont {Lu}, \citenamefont {Chu}, \citenamefont {Shen},\ and\
  \citenamefont {Niu}}]{Zhou_2008}%
  \BibitemOpen
  \bibfield  {author} {\bibinfo {author} {\bibfnamefont {B.}~\bibnamefont
  {Zhou}}, \bibinfo {author} {\bibfnamefont {H.-Z.}\ \bibnamefont {Lu}},
  \bibinfo {author} {\bibfnamefont {R.-L.}\ \bibnamefont {Chu}}, \bibinfo
  {author} {\bibfnamefont {S.-Q.}\ \bibnamefont {Shen}},\ and\ \bibinfo
  {author} {\bibfnamefont {Q.}~\bibnamefont {Niu}},\ }\bibfield  {title}
  {\bibinfo {title} {Finite size effects on helical edge states in a quantum
  spin-hall system},\ }\href {https://doi.org/10.1103/PhysRevLett.101.246807}
  {\bibfield  {journal} {\bibinfo  {journal} {Phys. Rev. Lett.}\ }\textbf
  {\bibinfo {volume} {101}},\ \bibinfo {pages} {246807} (\bibinfo {year}
  {2008})}\BibitemShut {NoStop}%
\bibitem [{\citenamefont {Potter}\ and\ \citenamefont
  {Lee}(2010)}]{Potter_Multi_2010}%
  \BibitemOpen
  \bibfield  {author} {\bibinfo {author} {\bibfnamefont {A.~C.}\ \bibnamefont
  {Potter}}\ and\ \bibinfo {author} {\bibfnamefont {P.~A.}\ \bibnamefont
  {Lee}},\ }\bibfield  {title} {\bibinfo {title} {Multichannel generalization
  of kitaev's majorana end states and a practical route to realize them in thin
  films},\ }\href {https://doi.org/10.1103/PhysRevLett.105.227003} {\bibfield
  {journal} {\bibinfo  {journal} {Phys. Rev. Lett.}\ }\textbf {\bibinfo
  {volume} {105}},\ \bibinfo {pages} {227003} (\bibinfo {year}
  {2010})}\BibitemShut {NoStop}%
\bibitem [{\citenamefont {Asmar}\ \emph {et~al.}(2018)\citenamefont {Asmar},
  \citenamefont {Sheehy},\ and\ \citenamefont {Vekhter}}]{Asmar_Topo_2018}%
  \BibitemOpen
  \bibfield  {author} {\bibinfo {author} {\bibfnamefont {M.~M.}\ \bibnamefont
  {Asmar}}, \bibinfo {author} {\bibfnamefont {D.~E.}\ \bibnamefont {Sheehy}},\
  and\ \bibinfo {author} {\bibfnamefont {I.}~\bibnamefont {Vekhter}},\
  }\bibfield  {title} {\bibinfo {title} {Topological phases of
  topological-insulator thin films},\ }\href
  {https://doi.org/10.1103/PhysRevB.97.075419} {\bibfield  {journal} {\bibinfo
  {journal} {Phys. Rev. B}\ }\textbf {\bibinfo {volume} {97}},\ \bibinfo
  {pages} {075419} (\bibinfo {year} {2018})}\BibitemShut {NoStop}%
\bibitem [{\citenamefont {He}\ \emph {et~al.}(2018)\citenamefont {He},
  \citenamefont {Yin}, \citenamefont {Yu}, \citenamefont {Grutter},
  \citenamefont {Pan}, \citenamefont {Chen}, \citenamefont {Che}, \citenamefont
  {Yu}, \citenamefont {Zhang}, \citenamefont {Shao}, \citenamefont {Stern},
  \citenamefont {Casas}, \citenamefont {Xia}, \citenamefont {Han},
  \citenamefont {Kirby}, \citenamefont {Lake}, \citenamefont {Law},\ and\
  \citenamefont {Wang}}]{He_Topo_2018}%
  \BibitemOpen
  \bibfield  {author} {\bibinfo {author} {\bibfnamefont {Q.~L.}\ \bibnamefont
  {He}}, \bibinfo {author} {\bibfnamefont {G.}~\bibnamefont {Yin}}, \bibinfo
  {author} {\bibfnamefont {L.}~\bibnamefont {Yu}}, \bibinfo {author}
  {\bibfnamefont {A.~J.}\ \bibnamefont {Grutter}}, \bibinfo {author}
  {\bibfnamefont {L.}~\bibnamefont {Pan}}, \bibinfo {author} {\bibfnamefont
  {C.-Z.}\ \bibnamefont {Chen}}, \bibinfo {author} {\bibfnamefont
  {X.}~\bibnamefont {Che}}, \bibinfo {author} {\bibfnamefont {G.}~\bibnamefont
  {Yu}}, \bibinfo {author} {\bibfnamefont {B.}~\bibnamefont {Zhang}}, \bibinfo
  {author} {\bibfnamefont {Q.}~\bibnamefont {Shao}}, \bibinfo {author}
  {\bibfnamefont {A.~L.}\ \bibnamefont {Stern}}, \bibinfo {author}
  {\bibfnamefont {B.}~\bibnamefont {Casas}}, \bibinfo {author} {\bibfnamefont
  {J.}~\bibnamefont {Xia}}, \bibinfo {author} {\bibfnamefont {X.}~\bibnamefont
  {Han}}, \bibinfo {author} {\bibfnamefont {B.~J.}\ \bibnamefont {Kirby}},
  \bibinfo {author} {\bibfnamefont {R.~K.}\ \bibnamefont {Lake}}, \bibinfo
  {author} {\bibfnamefont {K.~T.}\ \bibnamefont {Law}},\ and\ \bibinfo {author}
  {\bibfnamefont {K.~L.}\ \bibnamefont {Wang}},\ }\bibfield  {title} {\bibinfo
  {title} {Topological transitions induced by antiferromagnetism in a thin-film
  topological insulator},\ }\href
  {https://doi.org/10.1103/PhysRevLett.121.096802} {\bibfield  {journal}
  {\bibinfo  {journal} {Phys. Rev. Lett.}\ }\textbf {\bibinfo {volume} {121}},\
  \bibinfo {pages} {096802} (\bibinfo {year} {2018})}\BibitemShut {NoStop}%
\bibitem [{\citenamefont {Otrokov}\ \emph {et~al.}(2019)\citenamefont
  {Otrokov}, \citenamefont {Rusinov}, \citenamefont {Blanco-Rey}, \citenamefont
  {Hoffmann}, \citenamefont {Vyazovskaya}, \citenamefont {Eremeev},
  \citenamefont {Ernst}, \citenamefont {Echenique}, \citenamefont {Arnau},\
  and\ \citenamefont {Chulkov}}]{Otrokov_Thickness_2019}%
  \BibitemOpen
  \bibfield  {author} {\bibinfo {author} {\bibfnamefont {M.~M.}\ \bibnamefont
  {Otrokov}}, \bibinfo {author} {\bibfnamefont {I.~P.}\ \bibnamefont
  {Rusinov}}, \bibinfo {author} {\bibfnamefont {M.}~\bibnamefont {Blanco-Rey}},
  \bibinfo {author} {\bibfnamefont {M.}~\bibnamefont {Hoffmann}}, \bibinfo
  {author} {\bibfnamefont {A.~Y.}\ \bibnamefont {Vyazovskaya}}, \bibinfo
  {author} {\bibfnamefont {S.~V.}\ \bibnamefont {Eremeev}}, \bibinfo {author}
  {\bibfnamefont {A.}~\bibnamefont {Ernst}}, \bibinfo {author} {\bibfnamefont
  {P.~M.}\ \bibnamefont {Echenique}}, \bibinfo {author} {\bibfnamefont
  {A.}~\bibnamefont {Arnau}},\ and\ \bibinfo {author} {\bibfnamefont {E.~V.}\
  \bibnamefont {Chulkov}},\ }\bibfield  {title} {\bibinfo {title} {Unique
  thickness-dependent properties of the van der waals interlayer
  antiferromagnet ${\mathrm{mnbi}}_{2}{\mathrm{te}}_{4}$ films},\ }\href
  {https://doi.org/10.1103/PhysRevLett.122.107202} {\bibfield  {journal}
  {\bibinfo  {journal} {Phys. Rev. Lett.}\ }\textbf {\bibinfo {volume} {122}},\
  \bibinfo {pages} {107202} (\bibinfo {year} {2019})}\BibitemShut {NoStop}%
\bibitem [{\citenamefont {Chowdhury}\ \emph {et~al.}(2019)\citenamefont
  {Chowdhury}, \citenamefont {Garrity},\ and\ \citenamefont
  {Tavazza}}]{Chowdhury_Prediction_2019}%
  \BibitemOpen
  \bibfield  {author} {\bibinfo {author} {\bibfnamefont {S.}~\bibnamefont
  {Chowdhury}}, \bibinfo {author} {\bibfnamefont {K.~F.}\ \bibnamefont
  {Garrity}},\ and\ \bibinfo {author} {\bibfnamefont {F.}~\bibnamefont
  {Tavazza}},\ }\bibfield  {title} {\bibinfo {title} {Prediction of weyl
  semimetal and antiferromagnetic topological insulator phases in bi2mnse4},\
  }\href {https://doi.org/10.1038/s41524-019-0168-1} {\bibfield  {journal}
  {\bibinfo  {journal} {npj Computational Materials}\ }\textbf {\bibinfo
  {volume} {5}},\ \bibinfo {pages} {33} (\bibinfo {year} {2019})}\BibitemShut
  {NoStop}%
\bibitem [{\citenamefont {Li}\ \emph {et~al.}(2019)\citenamefont {Li},
  \citenamefont {Li}, \citenamefont {Du}, \citenamefont {Wang}, \citenamefont
  {Gu}, \citenamefont {Zhang}, \citenamefont {He}, \citenamefont {Duan},\ and\
  \citenamefont {Xu}}]{Li_Intrinsic_2019}%
  \BibitemOpen
  \bibfield  {author} {\bibinfo {author} {\bibfnamefont {J.}~\bibnamefont
  {Li}}, \bibinfo {author} {\bibfnamefont {Y.}~\bibnamefont {Li}}, \bibinfo
  {author} {\bibfnamefont {S.}~\bibnamefont {Du}}, \bibinfo {author}
  {\bibfnamefont {Z.}~\bibnamefont {Wang}}, \bibinfo {author} {\bibfnamefont
  {B.-L.}\ \bibnamefont {Gu}}, \bibinfo {author} {\bibfnamefont {S.-C.}\
  \bibnamefont {Zhang}}, \bibinfo {author} {\bibfnamefont {K.}~\bibnamefont
  {He}}, \bibinfo {author} {\bibfnamefont {W.}~\bibnamefont {Duan}},\ and\
  \bibinfo {author} {\bibfnamefont {Y.}~\bibnamefont {Xu}},\ }\bibfield
  {title} {\bibinfo {title} {Intrinsic magnetic topological insulators in van
  der waals layered mnbi<sub>2</sub>te<sub>4</sub>-family materials},\ }\href
  {https://doi.org/10.1126/sciadv.aaw5685} {\bibfield  {journal} {\bibinfo
  {journal} {Science Advances}\ }\textbf {\bibinfo {volume} {5}},\ \bibinfo
  {pages} {eaaw5685} (\bibinfo {year} {2019})},\ \Eprint
  {https://arxiv.org/abs/https://www.science.org/doi/pdf/10.1126/sciadv.aaw5685}
  {https://www.science.org/doi/pdf/10.1126/sciadv.aaw5685} \BibitemShut
  {NoStop}%
\bibitem [{\citenamefont {Lei}\ \emph {et~al.}(2020)\citenamefont {Lei},
  \citenamefont {Chen},\ and\ \citenamefont
  {MacDonald}}]{Chao_Magnetized_2020}%
  \BibitemOpen
  \bibfield  {author} {\bibinfo {author} {\bibfnamefont {C.}~\bibnamefont
  {Lei}}, \bibinfo {author} {\bibfnamefont {S.}~\bibnamefont {Chen}},\ and\
  \bibinfo {author} {\bibfnamefont {A.~H.}\ \bibnamefont {MacDonald}},\
  }\bibfield  {title} {\bibinfo {title} {Magnetized topological insulator
  multilayers},\ }\href {https://doi.org/10.1073/pnas.2014004117} {\bibfield
  {journal} {\bibinfo  {journal} {Proceedings of the National Academy of
  Sciences}\ }\textbf {\bibinfo {volume} {117}},\ \bibinfo {pages} {27224}
  (\bibinfo {year} {2020})},\ \Eprint
  {https://arxiv.org/abs/https://www.pnas.org/doi/pdf/10.1073/pnas.2014004117}
  {https://www.pnas.org/doi/pdf/10.1073/pnas.2014004117} \BibitemShut {NoStop}%
\bibitem [{\citenamefont {Liu}\ and\ \citenamefont
  {Hesjedal}(2023)}]{Liu_Magnteic_2023}%
  \BibitemOpen
  \bibfield  {author} {\bibinfo {author} {\bibfnamefont {J.}~\bibnamefont
  {Liu}}\ and\ \bibinfo {author} {\bibfnamefont {T.}~\bibnamefont {Hesjedal}},\
  }\bibfield  {title} {\bibinfo {title} {Magnetic topological insulator
  heterostructures: A review},\ }\href
  {https://doi.org/https://doi.org/10.1002/adma.202102427} {\bibfield
  {journal} {\bibinfo  {journal} {Advanced Materials}\ }\textbf {\bibinfo
  {volume} {35}},\ \bibinfo {pages} {2102427} (\bibinfo {year}
  {2023})}\BibitemShut {NoStop}%
\bibitem [{\citenamefont {Lygo}\ \emph {et~al.}(2023)\citenamefont {Lygo},
  \citenamefont {Guo}, \citenamefont {Rashidi}, \citenamefont {Huang},
  \citenamefont {Cuadros-Romero},\ and\ \citenamefont {Stemmer}}]{Lygo2023}%
  \BibitemOpen
  \bibfield  {author} {\bibinfo {author} {\bibfnamefont {A.~C.}\ \bibnamefont
  {Lygo}}, \bibinfo {author} {\bibfnamefont {B.}~\bibnamefont {Guo}}, \bibinfo
  {author} {\bibfnamefont {A.}~\bibnamefont {Rashidi}}, \bibinfo {author}
  {\bibfnamefont {V.}~\bibnamefont {Huang}}, \bibinfo {author} {\bibfnamefont
  {P.}~\bibnamefont {Cuadros-Romero}},\ and\ \bibinfo {author} {\bibfnamefont
  {S.}~\bibnamefont {Stemmer}},\ }\bibfield  {title} {\bibinfo {title}
  {Two-dimensional topological insulator state in cadmium arsenide thin
  films},\ }\href {https://doi.org/10.1103/PhysRevLett.130.046201} {\bibfield
  {journal} {\bibinfo  {journal} {Phys. Rev. Lett.}\ }\textbf {\bibinfo
  {volume} {130}},\ \bibinfo {pages} {046201} (\bibinfo {year}
  {2023})}\BibitemShut {NoStop}%
\bibitem [{\citenamefont {Guo}\ \emph {et~al.}(2023{\natexlab{a}})\citenamefont
  {Guo}, \citenamefont {Miao}, \citenamefont {Huang}, \citenamefont {Lygo},
  \citenamefont {Dai},\ and\ \citenamefont {Stemmer}}]{Guo2023}%
  \BibitemOpen
  \bibfield  {author} {\bibinfo {author} {\bibfnamefont {B.}~\bibnamefont
  {Guo}}, \bibinfo {author} {\bibfnamefont {W.}~\bibnamefont {Miao}}, \bibinfo
  {author} {\bibfnamefont {V.}~\bibnamefont {Huang}}, \bibinfo {author}
  {\bibfnamefont {A.~C.}\ \bibnamefont {Lygo}}, \bibinfo {author}
  {\bibfnamefont {X.}~\bibnamefont {Dai}},\ and\ \bibinfo {author}
  {\bibfnamefont {S.}~\bibnamefont {Stemmer}},\ }\bibfield  {title} {\bibinfo
  {title} {Zeeman field-induced two-dimensional weyl semimetal phase in cadmium
  arsenide},\ }\href {https://doi.org/10.1103/PhysRevLett.131.046601}
  {\bibfield  {journal} {\bibinfo  {journal} {Phys. Rev. Lett.}\ }\textbf
  {\bibinfo {volume} {131}},\ \bibinfo {pages} {046601} (\bibinfo {year}
  {2023}{\natexlab{a}})}\BibitemShut {NoStop}%
\bibitem [{\citenamefont {Lin}\ \emph {et~al.}(2023)\citenamefont {Lin},
  \citenamefont {Sun}, \citenamefont {Liu},\ and\ \citenamefont
  {Zhao}}]{Lin2023}%
  \BibitemOpen
  \bibfield  {author} {\bibinfo {author} {\bibfnamefont {H.-J.}\ \bibnamefont
  {Lin}}, \bibinfo {author} {\bibfnamefont {H.-P.}\ \bibnamefont {Sun}},
  \bibinfo {author} {\bibfnamefont {T.}~\bibnamefont {Liu}},\ and\ \bibinfo
  {author} {\bibfnamefont {P.-L.}\ \bibnamefont {Zhao}},\ }\bibfield  {title}
  {\bibinfo {title} {Tuning three-dimensional higher-order topological
  insulators by surface state hybridization},\ }\href
  {https://doi.org/10.1103/PhysRevB.108.165427} {\bibfield  {journal} {\bibinfo
   {journal} {Phys. Rev. B}\ }\textbf {\bibinfo {volume} {108}},\ \bibinfo
  {pages} {165427} (\bibinfo {year} {2023})}\BibitemShut {NoStop}%
\bibitem [{\citenamefont {Smith}\ \emph {et~al.}(2024)\citenamefont {Smith},
  \citenamefont {Quito}, \citenamefont {Burkov}, \citenamefont {Orth},\ and\
  \citenamefont {Martin}}]{Smith2024}%
  \BibitemOpen
  \bibfield  {author} {\bibinfo {author} {\bibfnamefont {M.}~\bibnamefont
  {Smith}}, \bibinfo {author} {\bibfnamefont {V.~L.}\ \bibnamefont {Quito}},
  \bibinfo {author} {\bibfnamefont {A.~A.}\ \bibnamefont {Burkov}}, \bibinfo
  {author} {\bibfnamefont {P.~P.}\ \bibnamefont {Orth}},\ and\ \bibinfo
  {author} {\bibfnamefont {I.}~\bibnamefont {Martin}},\ }\bibfield  {title}
  {\bibinfo {title} {Theory for ${\mathrm{cd}}_{3}{\mathrm{as}}_{2}$ thin films
  in the presence of magnetic fields},\ }\href
  {https://doi.org/10.1103/PhysRevB.109.155136} {\bibfield  {journal} {\bibinfo
   {journal} {Phys. Rev. B}\ }\textbf {\bibinfo {volume} {109}},\ \bibinfo
  {pages} {155136} (\bibinfo {year} {2024})}\BibitemShut {NoStop}%
\bibitem [{\citenamefont {Xiao}\ \emph {et~al.}(2015)\citenamefont {Xiao},
  \citenamefont {Yang}, \citenamefont {Liu}, \citenamefont {Li},\ and\
  \citenamefont {Zhou}}]{Xiao2015}%
  \BibitemOpen
  \bibfield  {author} {\bibinfo {author} {\bibfnamefont {X.}~\bibnamefont
  {Xiao}}, \bibinfo {author} {\bibfnamefont {S.~A.}\ \bibnamefont {Yang}},
  \bibinfo {author} {\bibfnamefont {Z.}~\bibnamefont {Liu}}, \bibinfo {author}
  {\bibfnamefont {H.}~\bibnamefont {Li}},\ and\ \bibinfo {author}
  {\bibfnamefont {G.}~\bibnamefont {Zhou}},\ }\bibfield  {title} {\bibinfo
  {title} {Anisotropic quantum confinement effect and electric control of
  surface states in dirac semimetal nanostructures},\ }\href
  {https://doi.org/10.1038/srep07898} {\bibfield  {journal} {\bibinfo
  {journal} {Scientific Reports}\ }\textbf {\bibinfo {volume} {5}},\ \bibinfo
  {pages} {7898} (\bibinfo {year} {2015})}\BibitemShut {NoStop}%
\bibitem [{\citenamefont {Cook}\ and\ \citenamefont
  {Nielsen}(2023)}]{Cook_FST_2023}%
  \BibitemOpen
  \bibfield  {author} {\bibinfo {author} {\bibfnamefont {A.~M.}\ \bibnamefont
  {Cook}}\ and\ \bibinfo {author} {\bibfnamefont {A.~E.~B.}\ \bibnamefont
  {Nielsen}},\ }\bibfield  {title} {\bibinfo {title} {Finite-size topology},\
  }\href {https://doi.org/10.1103/PhysRevB.108.045144} {\bibfield  {journal}
  {\bibinfo  {journal} {Phys. Rev. B}\ }\textbf {\bibinfo {volume} {108}},\
  \bibinfo {pages} {045144} (\bibinfo {year} {2023})}\BibitemShut {NoStop}%
\bibitem [{\citenamefont {Flores-Calderon}\ \emph {et~al.}(2023)\citenamefont
  {Flores-Calderon}, \citenamefont {Moessner},\ and\ \citenamefont
  {Cook}}]{Cook_TRI_2023}%
  \BibitemOpen
  \bibfield  {author} {\bibinfo {author} {\bibfnamefont {R.}~\bibnamefont
  {Flores-Calderon}}, \bibinfo {author} {\bibfnamefont {R.}~\bibnamefont
  {Moessner}},\ and\ \bibinfo {author} {\bibfnamefont {A.~M.}\ \bibnamefont
  {Cook}},\ }\bibfield  {title} {\bibinfo {title} {Time-reversal invariant
  finite-size topology},\ }\href {https://doi.org/10.1103/PhysRevB.108.125410}
  {\bibfield  {journal} {\bibinfo  {journal} {Phys. Rev. B}\ }\textbf {\bibinfo
  {volume} {108}},\ \bibinfo {pages} {125410} (\bibinfo {year}
  {2023})}\BibitemShut {NoStop}%
\bibitem [{\citenamefont {Pal}\ and\ \citenamefont
  {Cook}(2025)}]{Pal_FST_2025}%
  \BibitemOpen
  \bibfield  {author} {\bibinfo {author} {\bibfnamefont {A.}~\bibnamefont
  {Pal}}\ and\ \bibinfo {author} {\bibfnamefont {A.~M.}\ \bibnamefont {Cook}},\
  }\bibfield  {title} {\bibinfo {title} {Finite-size topological phases from
  semimetals},\ }\href {https://doi.org/10.1103/PhysRevB.111.035146} {\bibfield
   {journal} {\bibinfo  {journal} {Phys. Rev. B}\ }\textbf {\bibinfo {volume}
  {111}},\ \bibinfo {pages} {035146} (\bibinfo {year} {2025})}\BibitemShut
  {NoStop}%
\bibitem [{\citenamefont {Leis}\ \emph {et~al.}(2021)\citenamefont {Leis},
  \citenamefont {Schleenvoigt}, \citenamefont {Cherepanov}, \citenamefont
  {Lüpke}, \citenamefont {Schüffelgen}, \citenamefont {Mussler},
  \citenamefont {Grützmacher}, \citenamefont {Voigtländer},\ and\
  \citenamefont {Tautz}}]{Leis_Lifting_2021}%
  \BibitemOpen
  \bibfield  {author} {\bibinfo {author} {\bibfnamefont {A.}~\bibnamefont
  {Leis}}, \bibinfo {author} {\bibfnamefont {M.}~\bibnamefont {Schleenvoigt}},
  \bibinfo {author} {\bibfnamefont {V.}~\bibnamefont {Cherepanov}}, \bibinfo
  {author} {\bibfnamefont {F.}~\bibnamefont {Lüpke}}, \bibinfo {author}
  {\bibfnamefont {P.}~\bibnamefont {Schüffelgen}}, \bibinfo {author}
  {\bibfnamefont {G.}~\bibnamefont {Mussler}}, \bibinfo {author} {\bibfnamefont
  {D.}~\bibnamefont {Grützmacher}}, \bibinfo {author} {\bibfnamefont
  {B.}~\bibnamefont {Voigtländer}},\ and\ \bibinfo {author} {\bibfnamefont
  {F.~S.}\ \bibnamefont {Tautz}},\ }\bibfield  {title} {\bibinfo {title}
  {Lifting the spin-momentum locking in ultra-thin topological insulator
  films},\ }\href {https://doi.org/https://doi.org/10.1002/qute.202100083}
  {\bibfield  {journal} {\bibinfo  {journal} {Advanced Quantum Technologies}\
  }\textbf {\bibinfo {volume} {4}},\ \bibinfo {pages} {2100083} (\bibinfo
  {year} {2021})}\BibitemShut {NoStop}%
\bibitem [{\citenamefont {Zhang}\ \emph {et~al.}(2010)\citenamefont {Zhang},
  \citenamefont {He}, \citenamefont {Chang}, \citenamefont {Song},
  \citenamefont {Wang}, \citenamefont {Chen}, \citenamefont {Jia},
  \citenamefont {Fang}, \citenamefont {Dai}, \citenamefont {Shan},
  \citenamefont {Shen}, \citenamefont {Niu}, \citenamefont {Qi}, \citenamefont
  {Zhang}, \citenamefont {Ma},\ and\ \citenamefont
  {Xue}}]{Zhang_Crossover_2010}%
  \BibitemOpen
  \bibfield  {author} {\bibinfo {author} {\bibfnamefont {Y.}~\bibnamefont
  {Zhang}}, \bibinfo {author} {\bibfnamefont {K.}~\bibnamefont {He}}, \bibinfo
  {author} {\bibfnamefont {C.-Z.}\ \bibnamefont {Chang}}, \bibinfo {author}
  {\bibfnamefont {C.-L.}\ \bibnamefont {Song}}, \bibinfo {author}
  {\bibfnamefont {L.-L.}\ \bibnamefont {Wang}}, \bibinfo {author}
  {\bibfnamefont {X.}~\bibnamefont {Chen}}, \bibinfo {author} {\bibfnamefont
  {J.-F.}\ \bibnamefont {Jia}}, \bibinfo {author} {\bibfnamefont
  {Z.}~\bibnamefont {Fang}}, \bibinfo {author} {\bibfnamefont {X.}~\bibnamefont
  {Dai}}, \bibinfo {author} {\bibfnamefont {W.-Y.}\ \bibnamefont {Shan}},
  \bibinfo {author} {\bibfnamefont {S.-Q.}\ \bibnamefont {Shen}}, \bibinfo
  {author} {\bibfnamefont {Q.}~\bibnamefont {Niu}}, \bibinfo {author}
  {\bibfnamefont {X.-L.}\ \bibnamefont {Qi}}, \bibinfo {author} {\bibfnamefont
  {S.-C.}\ \bibnamefont {Zhang}}, \bibinfo {author} {\bibfnamefont {X.-C.}\
  \bibnamefont {Ma}},\ and\ \bibinfo {author} {\bibfnamefont {Q.-K.}\
  \bibnamefont {Xue}},\ }\bibfield  {title} {\bibinfo {title} {Crossover of the
  three-dimensional topological insulator bi2se3 to the two-dimensional
  limit},\ }\href {https://doi.org/10.1038/nphys1689} {\bibfield  {journal}
  {\bibinfo  {journal} {Nature Physics}\ }\textbf {\bibinfo {volume} {6}},\
  \bibinfo {pages} {584} (\bibinfo {year} {2010})}\BibitemShut {NoStop}%
\bibitem [{\citenamefont {Sakamoto}\ \emph {et~al.}(2010)\citenamefont
  {Sakamoto}, \citenamefont {Hirahara}, \citenamefont {Miyazaki}, \citenamefont
  {Kimura},\ and\ \citenamefont {Hasegawa}}]{Sakamoto_Spectro_2010}%
  \BibitemOpen
  \bibfield  {author} {\bibinfo {author} {\bibfnamefont {Y.}~\bibnamefont
  {Sakamoto}}, \bibinfo {author} {\bibfnamefont {T.}~\bibnamefont {Hirahara}},
  \bibinfo {author} {\bibfnamefont {H.}~\bibnamefont {Miyazaki}}, \bibinfo
  {author} {\bibfnamefont {S.-i.}\ \bibnamefont {Kimura}},\ and\ \bibinfo
  {author} {\bibfnamefont {S.}~\bibnamefont {Hasegawa}},\ }\bibfield  {title}
  {\bibinfo {title} {Spectroscopic evidence of a topological quantum phase
  transition in ultrathin ${\text{bi}}_{2}{\text{se}}_{3}$ films},\ }\href
  {https://doi.org/10.1103/PhysRevB.81.165432} {\bibfield  {journal} {\bibinfo
  {journal} {Phys. Rev. B}\ }\textbf {\bibinfo {volume} {81}},\ \bibinfo
  {pages} {165432} (\bibinfo {year} {2010})}\BibitemShut {NoStop}%
\bibitem [{\citenamefont {Geim}\ and\ \citenamefont
  {Grigorieva}(2013)}]{Geim2013}%
  \BibitemOpen
  \bibfield  {author} {\bibinfo {author} {\bibfnamefont {A.~K.}\ \bibnamefont
  {Geim}}\ and\ \bibinfo {author} {\bibfnamefont {I.~V.}\ \bibnamefont
  {Grigorieva}},\ }\bibfield  {title} {\bibinfo {title} {Van der waals
  heterostructures},\ }\href {https://doi.org/10.1038/nature12385} {\bibfield
  {journal} {\bibinfo  {journal} {Nature}\ }\textbf {\bibinfo {volume} {499}},\
  \bibinfo {pages} {419} (\bibinfo {year} {2013})}\BibitemShut {NoStop}%
\bibitem [{\citenamefont {Hu}\ \emph {et~al.}(2020)\citenamefont {Hu},
  \citenamefont {Gordon}, \citenamefont {Liu}, \citenamefont {Liu},
  \citenamefont {Zhou}, \citenamefont {Hao}, \citenamefont {Narayan},
  \citenamefont {Emmanouilidou}, \citenamefont {Sun}, \citenamefont {Liu},
  \citenamefont {Brawer}, \citenamefont {Ramirez}, \citenamefont {Ding},
  \citenamefont {Cao}, \citenamefont {Liu}, \citenamefont {Dessau},\ and\
  \citenamefont {Ni}}]{Hu2020}%
  \BibitemOpen
  \bibfield  {author} {\bibinfo {author} {\bibfnamefont {C.}~\bibnamefont
  {Hu}}, \bibinfo {author} {\bibfnamefont {K.~N.}\ \bibnamefont {Gordon}},
  \bibinfo {author} {\bibfnamefont {P.}~\bibnamefont {Liu}}, \bibinfo {author}
  {\bibfnamefont {J.}~\bibnamefont {Liu}}, \bibinfo {author} {\bibfnamefont
  {X.}~\bibnamefont {Zhou}}, \bibinfo {author} {\bibfnamefont {P.}~\bibnamefont
  {Hao}}, \bibinfo {author} {\bibfnamefont {D.}~\bibnamefont {Narayan}},
  \bibinfo {author} {\bibfnamefont {E.}~\bibnamefont {Emmanouilidou}}, \bibinfo
  {author} {\bibfnamefont {H.}~\bibnamefont {Sun}}, \bibinfo {author}
  {\bibfnamefont {Y.}~\bibnamefont {Liu}}, \bibinfo {author} {\bibfnamefont
  {H.}~\bibnamefont {Brawer}}, \bibinfo {author} {\bibfnamefont {A.~P.}\
  \bibnamefont {Ramirez}}, \bibinfo {author} {\bibfnamefont {L.}~\bibnamefont
  {Ding}}, \bibinfo {author} {\bibfnamefont {H.}~\bibnamefont {Cao}}, \bibinfo
  {author} {\bibfnamefont {Q.}~\bibnamefont {Liu}}, \bibinfo {author}
  {\bibfnamefont {D.}~\bibnamefont {Dessau}},\ and\ \bibinfo {author}
  {\bibfnamefont {N.}~\bibnamefont {Ni}},\ }\bibfield  {title} {\bibinfo
  {title} {A van der waals antiferromagnetic topological insulator with weak
  interlayer magnetic coupling},\ }\href
  {https://doi.org/10.1038/s41467-019-13814-x} {\bibfield  {journal} {\bibinfo
  {journal} {Nature Communications}\ }\textbf {\bibinfo {volume} {11}},\
  \bibinfo {pages} {97} (\bibinfo {year} {2020})}\BibitemShut {NoStop}%
\bibitem [{\citenamefont {Chong}\ \emph {et~al.}(2018)\citenamefont {Chong},
  \citenamefont {Han}, \citenamefont {Nagaoka}, \citenamefont {Tsuchikawa},
  \citenamefont {Liu}, \citenamefont {Liu}, \citenamefont {Vardeny},
  \citenamefont {Pesin}, \citenamefont {Lee}, \citenamefont {Sparks},\ and\
  \citenamefont {Deshpande}}]{Chong2018}%
  \BibitemOpen
  \bibfield  {author} {\bibinfo {author} {\bibfnamefont {S.~K.}\ \bibnamefont
  {Chong}}, \bibinfo {author} {\bibfnamefont {K.~B.}\ \bibnamefont {Han}},
  \bibinfo {author} {\bibfnamefont {A.}~\bibnamefont {Nagaoka}}, \bibinfo
  {author} {\bibfnamefont {R.}~\bibnamefont {Tsuchikawa}}, \bibinfo {author}
  {\bibfnamefont {R.}~\bibnamefont {Liu}}, \bibinfo {author} {\bibfnamefont
  {H.}~\bibnamefont {Liu}}, \bibinfo {author} {\bibfnamefont {Z.~V.}\
  \bibnamefont {Vardeny}}, \bibinfo {author} {\bibfnamefont {D.~A.}\
  \bibnamefont {Pesin}}, \bibinfo {author} {\bibfnamefont {C.}~\bibnamefont
  {Lee}}, \bibinfo {author} {\bibfnamefont {T.~D.}\ \bibnamefont {Sparks}},\
  and\ \bibinfo {author} {\bibfnamefont {V.~V.}\ \bibnamefont {Deshpande}},\
  }\bibfield  {title} {\bibinfo {title} {Topological insulator-based van der
  waals heterostructures for effective control of massless and massive dirac
  fermions},\ }\href {https://doi.org/10.1021/acs.nanolett.8b04291} {\bibfield
  {journal} {\bibinfo  {journal} {Nano Letters}\ }\textbf {\bibinfo {volume}
  {18}},\ \bibinfo {pages} {8047} (\bibinfo {year} {2018})}\BibitemShut
  {NoStop}%
\bibitem [{\citenamefont {Kou}\ \emph {et~al.}(2014)\citenamefont {Kou},
  \citenamefont {Wu}, \citenamefont {Felser}, \citenamefont {Frauenheim},
  \citenamefont {Chen},\ and\ \citenamefont {Yan}}]{Kou2014}%
  \BibitemOpen
  \bibfield  {author} {\bibinfo {author} {\bibfnamefont {L.}~\bibnamefont
  {Kou}}, \bibinfo {author} {\bibfnamefont {S.-C.}\ \bibnamefont {Wu}},
  \bibinfo {author} {\bibfnamefont {C.}~\bibnamefont {Felser}}, \bibinfo
  {author} {\bibfnamefont {T.}~\bibnamefont {Frauenheim}}, \bibinfo {author}
  {\bibfnamefont {C.}~\bibnamefont {Chen}},\ and\ \bibinfo {author}
  {\bibfnamefont {B.}~\bibnamefont {Yan}},\ }\bibfield  {title} {\bibinfo
  {title} {Robust 2d topological insulators in van der waals
  heterostructures},\ }\href {https://doi.org/10.1021/nn503789v} {\bibfield
  {journal} {\bibinfo  {journal} {ACS Nano}\ }\textbf {\bibinfo {volume} {8}},\
  \bibinfo {pages} {10448} (\bibinfo {year} {2014})}\BibitemShut {NoStop}%
\bibitem [{\citenamefont {Husain}\ \emph {et~al.}(2020)\citenamefont {Husain},
  \citenamefont {Gupta}, \citenamefont {Kumar}, \citenamefont {Kumar},
  \citenamefont {Behera}, \citenamefont {Brucas}, \citenamefont {Chaudhary},\
  and\ \citenamefont {Svedlindh}}]{Hussian2020}%
  \BibitemOpen
  \bibfield  {author} {\bibinfo {author} {\bibfnamefont {S.}~\bibnamefont
  {Husain}}, \bibinfo {author} {\bibfnamefont {R.}~\bibnamefont {Gupta}},
  \bibinfo {author} {\bibfnamefont {A.}~\bibnamefont {Kumar}}, \bibinfo
  {author} {\bibfnamefont {P.}~\bibnamefont {Kumar}}, \bibinfo {author}
  {\bibfnamefont {N.}~\bibnamefont {Behera}}, \bibinfo {author} {\bibfnamefont
  {R.}~\bibnamefont {Brucas}}, \bibinfo {author} {\bibfnamefont
  {S.}~\bibnamefont {Chaudhary}},\ and\ \bibinfo {author} {\bibfnamefont
  {P.}~\bibnamefont {Svedlindh}},\ }\bibfield  {title} {\bibinfo {title}
  {Emergence of spin–orbit torques in 2d transition metal dichalcogenides: A
  status update},\ }\href {https://doi.org/10.1063/5.0025318} {\bibfield
  {journal} {\bibinfo  {journal} {Applied Physics Reviews}\ }\textbf {\bibinfo
  {volume} {7}},\ \bibinfo {pages} {041312} (\bibinfo {year} {2020})},\ \Eprint
  {https://arxiv.org/abs/https://pubs.aip.org/aip/apr/article-pdf/doi/10.1063/5.0025318/13896109/041312\_1\_online.pdf}
  {https://pubs.aip.org/aip/apr/article-pdf/doi/10.1063/5.0025318/13896109/041312\_1\_online.pdf}
  \BibitemShut {NoStop}%
\bibitem [{\citenamefont {Burkov}\ \emph {et~al.}(2011)\citenamefont {Burkov},
  \citenamefont {Hook},\ and\ \citenamefont
  {Balents}}]{Burkov_Topological2011}%
  \BibitemOpen
  \bibfield  {author} {\bibinfo {author} {\bibfnamefont {A.~A.}\ \bibnamefont
  {Burkov}}, \bibinfo {author} {\bibfnamefont {M.~D.}\ \bibnamefont {Hook}},\
  and\ \bibinfo {author} {\bibfnamefont {L.}~\bibnamefont {Balents}},\
  }\bibfield  {title} {\bibinfo {title} {Topological nodal semimetals},\ }\href
  {https://doi.org/10.1103/PhysRevB.84.235126} {\bibfield  {journal} {\bibinfo
  {journal} {Phys. Rev. B}\ }\textbf {\bibinfo {volume} {84}},\ \bibinfo
  {pages} {235126} (\bibinfo {year} {2011})}\BibitemShut {NoStop}%
\bibitem [{\citenamefont {Schnyder}\ and\ \citenamefont
  {Ryu}(2011)}]{Schnyder_Ryu_2011}%
  \BibitemOpen
  \bibfield  {author} {\bibinfo {author} {\bibfnamefont {A.~P.}\ \bibnamefont
  {Schnyder}}\ and\ \bibinfo {author} {\bibfnamefont {S.}~\bibnamefont {Ryu}},\
  }\bibfield  {title} {\bibinfo {title} {Topological phases and surface flat
  bands in superconductors without inversion symmetry},\ }\href
  {https://doi.org/10.1103/PhysRevB.84.060504} {\bibfield  {journal} {\bibinfo
  {journal} {Phys. Rev. B}\ }\textbf {\bibinfo {volume} {84}},\ \bibinfo
  {pages} {060504} (\bibinfo {year} {2011})}\BibitemShut {NoStop}%
\bibitem [{\citenamefont {Chiu}\ and\ \citenamefont
  {Schnyder}(2014)}]{Chiu_Classification2014}%
  \BibitemOpen
  \bibfield  {author} {\bibinfo {author} {\bibfnamefont {C.-K.}\ \bibnamefont
  {Chiu}}\ and\ \bibinfo {author} {\bibfnamefont {A.~P.}\ \bibnamefont
  {Schnyder}},\ }\bibfield  {title} {\bibinfo {title} {Classification of
  reflection-symmetry-protected topological semimetals and nodal
  superconductors},\ }\href {https://doi.org/10.1103/PhysRevB.90.205136}
  {\bibfield  {journal} {\bibinfo  {journal} {Phys. Rev. B}\ }\textbf {\bibinfo
  {volume} {90}},\ \bibinfo {pages} {205136} (\bibinfo {year}
  {2014})}\BibitemShut {NoStop}%
\bibitem [{\citenamefont {Fang}\ \emph {et~al.}(2015)\citenamefont {Fang},
  \citenamefont {Chen}, \citenamefont {Kee},\ and\ \citenamefont
  {Fu}}]{Fang_Topological2015}%
  \BibitemOpen
  \bibfield  {author} {\bibinfo {author} {\bibfnamefont {C.}~\bibnamefont
  {Fang}}, \bibinfo {author} {\bibfnamefont {Y.}~\bibnamefont {Chen}}, \bibinfo
  {author} {\bibfnamefont {H.-Y.}\ \bibnamefont {Kee}},\ and\ \bibinfo {author}
  {\bibfnamefont {L.}~\bibnamefont {Fu}},\ }\bibfield  {title} {\bibinfo
  {title} {Topological nodal line semimetals with and without spin-orbital
  coupling},\ }\href {https://doi.org/10.1103/PhysRevB.92.081201} {\bibfield
  {journal} {\bibinfo  {journal} {Phys. Rev. B}\ }\textbf {\bibinfo {volume}
  {92}},\ \bibinfo {pages} {081201} (\bibinfo {year} {2015})}\BibitemShut
  {NoStop}%
\bibitem [{\citenamefont {Chen}\ \emph {et~al.}(2015)\citenamefont {Chen},
  \citenamefont {Xie}, \citenamefont {Yang}, \citenamefont {Pan}, \citenamefont
  {Zhang}, \citenamefont {Cohen},\ and\ \citenamefont {Zhang}}]{Chen2015}%
  \BibitemOpen
  \bibfield  {author} {\bibinfo {author} {\bibfnamefont {Y.}~\bibnamefont
  {Chen}}, \bibinfo {author} {\bibfnamefont {Y.}~\bibnamefont {Xie}}, \bibinfo
  {author} {\bibfnamefont {S.~A.}\ \bibnamefont {Yang}}, \bibinfo {author}
  {\bibfnamefont {H.}~\bibnamefont {Pan}}, \bibinfo {author} {\bibfnamefont
  {F.}~\bibnamefont {Zhang}}, \bibinfo {author} {\bibfnamefont {M.~L.}\
  \bibnamefont {Cohen}},\ and\ \bibinfo {author} {\bibfnamefont
  {S.}~\bibnamefont {Zhang}},\ }\bibfield  {title} {\bibinfo {title}
  {Nanostructured carbon allotropes with weyl-like loops and points},\ }\href
  {https://doi.org/10.1021/acs.nanolett.5b02978} {\bibfield  {journal}
  {\bibinfo  {journal} {Nano Letters}\ }\textbf {\bibinfo {volume} {15}},\
  \bibinfo {pages} {6974} (\bibinfo {year} {2015})}\BibitemShut {NoStop}%
\bibitem [{\citenamefont {Bian}\ \emph {et~al.}(2016)\citenamefont {Bian},
  \citenamefont {Chang}, \citenamefont {Sankar}, \citenamefont {Xu},
  \citenamefont {Zheng}, \citenamefont {Neupert}, \citenamefont {Chiu},
  \citenamefont {Huang}, \citenamefont {Chang}, \citenamefont {Belopolski},
  \citenamefont {Sanchez}, \citenamefont {Neupane}, \citenamefont {Alidoust},
  \citenamefont {Liu}, \citenamefont {Wang}, \citenamefont {Lee}, \citenamefont
  {Jeng}, \citenamefont {Zhang}, \citenamefont {Yuan}, \citenamefont {Jia},
  \citenamefont {Bansil}, \citenamefont {Chou}, \citenamefont {Lin},\ and\
  \citenamefont {Hasan}}]{Bian_Topological_2016}%
  \BibitemOpen
  \bibfield  {author} {\bibinfo {author} {\bibfnamefont {G.}~\bibnamefont
  {Bian}}, \bibinfo {author} {\bibfnamefont {T.-R.}\ \bibnamefont {Chang}},
  \bibinfo {author} {\bibfnamefont {R.}~\bibnamefont {Sankar}}, \bibinfo
  {author} {\bibfnamefont {S.-Y.}\ \bibnamefont {Xu}}, \bibinfo {author}
  {\bibfnamefont {H.}~\bibnamefont {Zheng}}, \bibinfo {author} {\bibfnamefont
  {T.}~\bibnamefont {Neupert}}, \bibinfo {author} {\bibfnamefont {C.-K.}\
  \bibnamefont {Chiu}}, \bibinfo {author} {\bibfnamefont {S.-M.}\ \bibnamefont
  {Huang}}, \bibinfo {author} {\bibfnamefont {G.}~\bibnamefont {Chang}},
  \bibinfo {author} {\bibfnamefont {I.}~\bibnamefont {Belopolski}}, \bibinfo
  {author} {\bibfnamefont {D.~S.}\ \bibnamefont {Sanchez}}, \bibinfo {author}
  {\bibfnamefont {M.}~\bibnamefont {Neupane}}, \bibinfo {author} {\bibfnamefont
  {N.}~\bibnamefont {Alidoust}}, \bibinfo {author} {\bibfnamefont
  {C.}~\bibnamefont {Liu}}, \bibinfo {author} {\bibfnamefont {B.}~\bibnamefont
  {Wang}}, \bibinfo {author} {\bibfnamefont {C.-C.}\ \bibnamefont {Lee}},
  \bibinfo {author} {\bibfnamefont {H.-T.}\ \bibnamefont {Jeng}}, \bibinfo
  {author} {\bibfnamefont {C.}~\bibnamefont {Zhang}}, \bibinfo {author}
  {\bibfnamefont {Z.}~\bibnamefont {Yuan}}, \bibinfo {author} {\bibfnamefont
  {S.}~\bibnamefont {Jia}}, \bibinfo {author} {\bibfnamefont {A.}~\bibnamefont
  {Bansil}}, \bibinfo {author} {\bibfnamefont {F.}~\bibnamefont {Chou}},
  \bibinfo {author} {\bibfnamefont {H.}~\bibnamefont {Lin}},\ and\ \bibinfo
  {author} {\bibfnamefont {M.~Z.}\ \bibnamefont {Hasan}},\ }\bibfield  {title}
  {\bibinfo {title} {Topological nodal-line fermions in spin-orbit metal
  pbtase2},\ }\href {https://doi.org/10.1038/ncomms10556} {\bibfield  {journal}
  {\bibinfo  {journal} {Nature Communications}\ }\textbf {\bibinfo {volume}
  {7}},\ \bibinfo {pages} {10556} (\bibinfo {year} {2016})}\BibitemShut
  {NoStop}%
\bibitem [{\citenamefont {Neupane}\ \emph {et~al.}(2016)\citenamefont
  {Neupane}, \citenamefont {Belopolski}, \citenamefont {Hosen}, \citenamefont
  {Sanchez}, \citenamefont {Sankar}, \citenamefont {Szlawska}, \citenamefont
  {Xu}, \citenamefont {Dimitri}, \citenamefont {Dhakal}, \citenamefont
  {Maldonado}, \citenamefont {Oppeneer}, \citenamefont {Kaczorowski},
  \citenamefont {Chou}, \citenamefont {Hasan},\ and\ \citenamefont
  {Durakiewicz}}]{Neupane_Observation_2016}%
  \BibitemOpen
  \bibfield  {author} {\bibinfo {author} {\bibfnamefont {M.}~\bibnamefont
  {Neupane}}, \bibinfo {author} {\bibfnamefont {I.}~\bibnamefont {Belopolski}},
  \bibinfo {author} {\bibfnamefont {M.~M.}\ \bibnamefont {Hosen}}, \bibinfo
  {author} {\bibfnamefont {D.~S.}\ \bibnamefont {Sanchez}}, \bibinfo {author}
  {\bibfnamefont {R.}~\bibnamefont {Sankar}}, \bibinfo {author} {\bibfnamefont
  {M.}~\bibnamefont {Szlawska}}, \bibinfo {author} {\bibfnamefont {S.-Y.}\
  \bibnamefont {Xu}}, \bibinfo {author} {\bibfnamefont {K.}~\bibnamefont
  {Dimitri}}, \bibinfo {author} {\bibfnamefont {N.}~\bibnamefont {Dhakal}},
  \bibinfo {author} {\bibfnamefont {P.}~\bibnamefont {Maldonado}}, \bibinfo
  {author} {\bibfnamefont {P.~M.}\ \bibnamefont {Oppeneer}}, \bibinfo {author}
  {\bibfnamefont {D.}~\bibnamefont {Kaczorowski}}, \bibinfo {author}
  {\bibfnamefont {F.}~\bibnamefont {Chou}}, \bibinfo {author} {\bibfnamefont
  {M.~Z.}\ \bibnamefont {Hasan}},\ and\ \bibinfo {author} {\bibfnamefont
  {T.}~\bibnamefont {Durakiewicz}},\ }\bibfield  {title} {\bibinfo {title}
  {Observation of topological nodal fermion semimetal phase in zrsis},\ }\href
  {https://doi.org/10.1103/PhysRevB.93.201104} {\bibfield  {journal} {\bibinfo
  {journal} {Phys. Rev. B}\ }\textbf {\bibinfo {volume} {93}},\ \bibinfo
  {pages} {201104} (\bibinfo {year} {2016})}\BibitemShut {NoStop}%
\bibitem [{\citenamefont {Bzdu{\v{s}}ek}\ \emph {et~al.}(2016)\citenamefont
  {Bzdu{\v{s}}ek}, \citenamefont {Wu}, \citenamefont {R{\"u}egg}, \citenamefont
  {Sigrist},\ and\ \citenamefont {Soluyanov}}]{Bzdusek2016}%
  \BibitemOpen
  \bibfield  {author} {\bibinfo {author} {\bibfnamefont {T.}~\bibnamefont
  {Bzdu{\v{s}}ek}}, \bibinfo {author} {\bibfnamefont {Q.}~\bibnamefont {Wu}},
  \bibinfo {author} {\bibfnamefont {A.}~\bibnamefont {R{\"u}egg}}, \bibinfo
  {author} {\bibfnamefont {M.}~\bibnamefont {Sigrist}},\ and\ \bibinfo {author}
  {\bibfnamefont {A.~A.}\ \bibnamefont {Soluyanov}},\ }\bibfield  {title}
  {\bibinfo {title} {Nodal-chain metals},\ }\href
  {https://doi.org/10.1038/nature19099} {\bibfield  {journal} {\bibinfo
  {journal} {Nature}\ }\textbf {\bibinfo {volume} {538}},\ \bibinfo {pages}
  {75} (\bibinfo {year} {2016})}\BibitemShut {NoStop}%
\bibitem [{\citenamefont {Bzdu\ifmmode~\check{s}\else \v{s}\fi{}ek}\ and\
  \citenamefont {Sigrist}(2017)}]{Bzdusek2017}%
  \BibitemOpen
  \bibfield  {author} {\bibinfo {author} {\bibfnamefont {T.~c.~v.}\
  \bibnamefont {Bzdu\ifmmode~\check{s}\else \v{s}\fi{}ek}}\ and\ \bibinfo
  {author} {\bibfnamefont {M.}~\bibnamefont {Sigrist}},\ }\bibfield  {title}
  {\bibinfo {title} {Robust doubly charged nodal lines and nodal surfaces in
  centrosymmetric systems},\ }\href
  {https://doi.org/10.1103/PhysRevB.96.155105} {\bibfield  {journal} {\bibinfo
  {journal} {Phys. Rev. B}\ }\textbf {\bibinfo {volume} {96}},\ \bibinfo
  {pages} {155105} (\bibinfo {year} {2017})}\BibitemShut {NoStop}%
\bibitem [{\citenamefont {Schoop}\ \emph {et~al.}(2016)\citenamefont {Schoop},
  \citenamefont {Ali}, \citenamefont {Stra{\ss}er}, \citenamefont {Topp},
  \citenamefont {Varykhalov}, \citenamefont {Marchenko}, \citenamefont
  {Duppel}, \citenamefont {Parkin}, \citenamefont {Lotsch},\ and\ \citenamefont
  {Ast}}]{Schoop_DiraCone_2016}%
  \BibitemOpen
  \bibfield  {author} {\bibinfo {author} {\bibfnamefont {L.~M.}\ \bibnamefont
  {Schoop}}, \bibinfo {author} {\bibfnamefont {M.~N.}\ \bibnamefont {Ali}},
  \bibinfo {author} {\bibfnamefont {C.}~\bibnamefont {Stra{\ss}er}}, \bibinfo
  {author} {\bibfnamefont {A.}~\bibnamefont {Topp}}, \bibinfo {author}
  {\bibfnamefont {A.}~\bibnamefont {Varykhalov}}, \bibinfo {author}
  {\bibfnamefont {D.}~\bibnamefont {Marchenko}}, \bibinfo {author}
  {\bibfnamefont {V.}~\bibnamefont {Duppel}}, \bibinfo {author} {\bibfnamefont
  {S.~S.~P.}\ \bibnamefont {Parkin}}, \bibinfo {author} {\bibfnamefont {B.~V.}\
  \bibnamefont {Lotsch}},\ and\ \bibinfo {author} {\bibfnamefont {C.~R.}\
  \bibnamefont {Ast}},\ }\bibfield  {title} {\bibinfo {title} {Dirac cone
  protected by non-symmorphic symmetry and three-dimensional dirac line node in
  zrsis},\ }\href {https://doi.org/10.1038/ncomms11696} {\bibfield  {journal}
  {\bibinfo  {journal} {Nature Communications}\ }\textbf {\bibinfo {volume}
  {7}},\ \bibinfo {pages} {11696} (\bibinfo {year} {2016})}\BibitemShut
  {NoStop}%
\bibitem [{\citenamefont {Wang}\ \emph {et~al.}(2017)\citenamefont {Wang},
  \citenamefont {Ma}, \citenamefont {Emmanouilidou}, \citenamefont {Shen},
  \citenamefont {Hsu}, \citenamefont {Zhou}, \citenamefont {Zuo}, \citenamefont
  {Song}, \citenamefont {Xu}, \citenamefont {Wang}, \citenamefont {Huang},
  \citenamefont {Ni},\ and\ \citenamefont {Liu}}]{Wang2017}%
  \BibitemOpen
  \bibfield  {author} {\bibinfo {author} {\bibfnamefont {X.-B.}\ \bibnamefont
  {Wang}}, \bibinfo {author} {\bibfnamefont {X.-M.}\ \bibnamefont {Ma}},
  \bibinfo {author} {\bibfnamefont {E.}~\bibnamefont {Emmanouilidou}}, \bibinfo
  {author} {\bibfnamefont {B.}~\bibnamefont {Shen}}, \bibinfo {author}
  {\bibfnamefont {C.-H.}\ \bibnamefont {Hsu}}, \bibinfo {author} {\bibfnamefont
  {C.-S.}\ \bibnamefont {Zhou}}, \bibinfo {author} {\bibfnamefont
  {Y.}~\bibnamefont {Zuo}}, \bibinfo {author} {\bibfnamefont {R.-R.}\
  \bibnamefont {Song}}, \bibinfo {author} {\bibfnamefont {S.-Y.}\ \bibnamefont
  {Xu}}, \bibinfo {author} {\bibfnamefont {G.}~\bibnamefont {Wang}}, \bibinfo
  {author} {\bibfnamefont {L.}~\bibnamefont {Huang}}, \bibinfo {author}
  {\bibfnamefont {N.}~\bibnamefont {Ni}},\ and\ \bibinfo {author}
  {\bibfnamefont {C.}~\bibnamefont {Liu}},\ }\bibfield  {title} {\bibinfo
  {title} {Topological surface electronic states in candidate nodal-line
  semimetal caagas},\ }\href {https://doi.org/10.1103/PhysRevB.96.161112}
  {\bibfield  {journal} {\bibinfo  {journal} {Phys. Rev. B}\ }\textbf {\bibinfo
  {volume} {96}},\ \bibinfo {pages} {161112} (\bibinfo {year}
  {2017})}\BibitemShut {NoStop}%
\bibitem [{\citenamefont {Lou}\ \emph {et~al.}(2018)\citenamefont {Lou},
  \citenamefont {Guo}, \citenamefont {Li}, \citenamefont {Wang}, \citenamefont
  {Liu}, \citenamefont {Sun}, \citenamefont {Li}, \citenamefont {Wu},
  \citenamefont {Wang}, \citenamefont {Sun}, \citenamefont {Shen},
  \citenamefont {Huang}, \citenamefont {Liu}, \citenamefont {Lu}, \citenamefont
  {Lei}, \citenamefont {Ding},\ and\ \citenamefont {Wang}}]{Lou2018}%
  \BibitemOpen
  \bibfield  {author} {\bibinfo {author} {\bibfnamefont {R.}~\bibnamefont
  {Lou}}, \bibinfo {author} {\bibfnamefont {P.}~\bibnamefont {Guo}}, \bibinfo
  {author} {\bibfnamefont {M.}~\bibnamefont {Li}}, \bibinfo {author}
  {\bibfnamefont {Q.}~\bibnamefont {Wang}}, \bibinfo {author} {\bibfnamefont
  {Z.}~\bibnamefont {Liu}}, \bibinfo {author} {\bibfnamefont {S.}~\bibnamefont
  {Sun}}, \bibinfo {author} {\bibfnamefont {C.}~\bibnamefont {Li}}, \bibinfo
  {author} {\bibfnamefont {X.}~\bibnamefont {Wu}}, \bibinfo {author}
  {\bibfnamefont {Z.}~\bibnamefont {Wang}}, \bibinfo {author} {\bibfnamefont
  {Z.}~\bibnamefont {Sun}}, \bibinfo {author} {\bibfnamefont {D.}~\bibnamefont
  {Shen}}, \bibinfo {author} {\bibfnamefont {Y.}~\bibnamefont {Huang}},
  \bibinfo {author} {\bibfnamefont {K.}~\bibnamefont {Liu}}, \bibinfo {author}
  {\bibfnamefont {Z.-Y.}\ \bibnamefont {Lu}}, \bibinfo {author} {\bibfnamefont
  {H.}~\bibnamefont {Lei}}, \bibinfo {author} {\bibfnamefont {H.}~\bibnamefont
  {Ding}},\ and\ \bibinfo {author} {\bibfnamefont {S.}~\bibnamefont {Wang}},\
  }\bibfield  {title} {\bibinfo {title} {Experimental observation of bulk nodal
  lines and electronic surface states in {ZrB}2},\ }\bibfield  {journal}
  {\bibinfo  {journal} {npj Quantum Materials}\ }\textbf {\bibinfo {volume}
  {3}},\ \href {https://doi.org/10.1038/s41535-018-0121-4}
  {10.1038/s41535-018-0121-4} (\bibinfo {year} {2018})\BibitemShut {NoStop}%
\bibitem [{\citenamefont {Belopolski}\ \emph {et~al.}(2019)\citenamefont
  {Belopolski}, \citenamefont {Manna}, \citenamefont {Sanchez}, \citenamefont
  {Chang}, \citenamefont {Ernst}, \citenamefont {Yin}, \citenamefont {Zhang},
  \citenamefont {Cochran}, \citenamefont {Shumiya}, \citenamefont {Zheng},
  \citenamefont {Singh}, \citenamefont {Bian}, \citenamefont {Multer},
  \citenamefont {Litskevich}, \citenamefont {Zhou}, \citenamefont {Huang},
  \citenamefont {Wang}, \citenamefont {Chang}, \citenamefont {Xu},
  \citenamefont {Bansil}, \citenamefont {Felser}, \citenamefont {Lin},\ and\
  \citenamefont {Hasan}}]{Belopolski2019}%
  \BibitemOpen
  \bibfield  {author} {\bibinfo {author} {\bibfnamefont {I.}~\bibnamefont
  {Belopolski}}, \bibinfo {author} {\bibfnamefont {K.}~\bibnamefont {Manna}},
  \bibinfo {author} {\bibfnamefont {D.~S.}\ \bibnamefont {Sanchez}}, \bibinfo
  {author} {\bibfnamefont {G.}~\bibnamefont {Chang}}, \bibinfo {author}
  {\bibfnamefont {B.}~\bibnamefont {Ernst}}, \bibinfo {author} {\bibfnamefont
  {J.}~\bibnamefont {Yin}}, \bibinfo {author} {\bibfnamefont {S.~S.}\
  \bibnamefont {Zhang}}, \bibinfo {author} {\bibfnamefont {T.}~\bibnamefont
  {Cochran}}, \bibinfo {author} {\bibfnamefont {N.}~\bibnamefont {Shumiya}},
  \bibinfo {author} {\bibfnamefont {H.}~\bibnamefont {Zheng}}, \bibinfo
  {author} {\bibfnamefont {B.}~\bibnamefont {Singh}}, \bibinfo {author}
  {\bibfnamefont {G.}~\bibnamefont {Bian}}, \bibinfo {author} {\bibfnamefont
  {D.}~\bibnamefont {Multer}}, \bibinfo {author} {\bibfnamefont
  {M.}~\bibnamefont {Litskevich}}, \bibinfo {author} {\bibfnamefont
  {X.}~\bibnamefont {Zhou}}, \bibinfo {author} {\bibfnamefont {S.-M.}\
  \bibnamefont {Huang}}, \bibinfo {author} {\bibfnamefont {B.}~\bibnamefont
  {Wang}}, \bibinfo {author} {\bibfnamefont {T.-R.}\ \bibnamefont {Chang}},
  \bibinfo {author} {\bibfnamefont {S.-Y.}\ \bibnamefont {Xu}}, \bibinfo
  {author} {\bibfnamefont {A.}~\bibnamefont {Bansil}}, \bibinfo {author}
  {\bibfnamefont {C.}~\bibnamefont {Felser}}, \bibinfo {author} {\bibfnamefont
  {H.}~\bibnamefont {Lin}},\ and\ \bibinfo {author} {\bibfnamefont {M.~Z.}\
  \bibnamefont {Hasan}},\ }\bibfield  {title} {\bibinfo {title} {Discovery of
  topological weyl fermion lines and drumhead surface states in a room
  temperature magnet},\ }\href {https://doi.org/10.1126/science.aav2327}
  {\bibfield  {journal} {\bibinfo  {journal} {Science}\ }\textbf {\bibinfo
  {volume} {365}},\ \bibinfo {pages} {1278} (\bibinfo {year}
  {2019})}\BibitemShut {NoStop}%
\bibitem [{\citenamefont {Hosen}\ \emph {et~al.}(2020)\citenamefont {Hosen},
  \citenamefont {Dhakal}, \citenamefont {Wang}, \citenamefont {Poudel},
  \citenamefont {Dimitri}, \citenamefont {Kabir}, \citenamefont {Sims},
  \citenamefont {Regmi}, \citenamefont {Gofryk}, \citenamefont {Kaczorowski},
  \citenamefont {Bansil},\ and\ \citenamefont {Neupane}}]{Hosen_2020}%
  \BibitemOpen
  \bibfield  {author} {\bibinfo {author} {\bibfnamefont {M.~M.}\ \bibnamefont
  {Hosen}}, \bibinfo {author} {\bibfnamefont {G.}~\bibnamefont {Dhakal}},
  \bibinfo {author} {\bibfnamefont {B.}~\bibnamefont {Wang}}, \bibinfo {author}
  {\bibfnamefont {N.}~\bibnamefont {Poudel}}, \bibinfo {author} {\bibfnamefont
  {K.}~\bibnamefont {Dimitri}}, \bibinfo {author} {\bibfnamefont
  {F.}~\bibnamefont {Kabir}}, \bibinfo {author} {\bibfnamefont
  {C.}~\bibnamefont {Sims}}, \bibinfo {author} {\bibfnamefont {S.}~\bibnamefont
  {Regmi}}, \bibinfo {author} {\bibfnamefont {K.}~\bibnamefont {Gofryk}},
  \bibinfo {author} {\bibfnamefont {D.}~\bibnamefont {Kaczorowski}}, \bibinfo
  {author} {\bibfnamefont {A.}~\bibnamefont {Bansil}},\ and\ \bibinfo {author}
  {\bibfnamefont {M.}~\bibnamefont {Neupane}},\ }\bibfield  {title} {\bibinfo
  {title} {Experimental observation of drumhead surface states in {SrAs}3},\
  }\bibfield  {journal} {\bibinfo  {journal} {Scientific Reports}\ }\textbf
  {\bibinfo {volume} {10}},\ \href {https://doi.org/10.1038/s41598-020-59200-2}
  {10.1038/s41598-020-59200-2} (\bibinfo {year} {2020})\BibitemShut {NoStop}%
\bibitem [{\citenamefont {Chen}\ \emph {et~al.}(2021)\citenamefont {Chen},
  \citenamefont {Liu}, \citenamefont {Yang}, \citenamefont {Chen},
  \citenamefont {Liu}, \citenamefont {Huang}, \citenamefont {Zhang},
  \citenamefont {Zhang}, \citenamefont {Liu},\ and\ \citenamefont
  {Shen}}]{Chen2021}%
  \BibitemOpen
  \bibfield  {author} {\bibinfo {author} {\bibfnamefont {W.}~\bibnamefont
  {Chen}}, \bibinfo {author} {\bibfnamefont {L.}~\bibnamefont {Liu}}, \bibinfo
  {author} {\bibfnamefont {W.}~\bibnamefont {Yang}}, \bibinfo {author}
  {\bibfnamefont {D.}~\bibnamefont {Chen}}, \bibinfo {author} {\bibfnamefont
  {Z.}~\bibnamefont {Liu}}, \bibinfo {author} {\bibfnamefont {Y.}~\bibnamefont
  {Huang}}, \bibinfo {author} {\bibfnamefont {T.}~\bibnamefont {Zhang}},
  \bibinfo {author} {\bibfnamefont {H.}~\bibnamefont {Zhang}}, \bibinfo
  {author} {\bibfnamefont {Z.}~\bibnamefont {Liu}},\ and\ \bibinfo {author}
  {\bibfnamefont {D.~W.}\ \bibnamefont {Shen}},\ }\bibfield  {title} {\bibinfo
  {title} {Evidence of topological nodal lines and surface states in the
  centrosymmetric superconductor ${\mathrm{sntas}}_{2}$},\ }\href
  {https://doi.org/10.1103/PhysRevB.103.035133} {\bibfield  {journal} {\bibinfo
   {journal} {Phys. Rev. B}\ }\textbf {\bibinfo {volume} {103}},\ \bibinfo
  {pages} {035133} (\bibinfo {year} {2021})}\BibitemShut {NoStop}%
\bibitem [{\citenamefont {Gao}\ \emph {et~al.}(2023)\citenamefont {Gao},
  \citenamefont {Zhu}, \citenamefont {Chen}, \citenamefont {Liang},
  \citenamefont {Wu}, \citenamefont {Zhu}, \citenamefont {Han}, \citenamefont
  {Li}, \citenamefont {Liu}, \citenamefont {Zheng}, \citenamefont {Lu},\ and\
  \citenamefont {Tian}}]{Gao2023}%
  \BibitemOpen
  \bibfield  {author} {\bibinfo {author} {\bibfnamefont {W.}~\bibnamefont
  {Gao}}, \bibinfo {author} {\bibfnamefont {M.}~\bibnamefont {Zhu}}, \bibinfo
  {author} {\bibfnamefont {D.}~\bibnamefont {Chen}}, \bibinfo {author}
  {\bibfnamefont {X.}~\bibnamefont {Liang}}, \bibinfo {author} {\bibfnamefont
  {Y.}~\bibnamefont {Wu}}, \bibinfo {author} {\bibfnamefont {A.}~\bibnamefont
  {Zhu}}, \bibinfo {author} {\bibfnamefont {Y.}~\bibnamefont {Han}}, \bibinfo
  {author} {\bibfnamefont {L.}~\bibnamefont {Li}}, \bibinfo {author}
  {\bibfnamefont {X.}~\bibnamefont {Liu}}, \bibinfo {author} {\bibfnamefont
  {G.}~\bibnamefont {Zheng}}, \bibinfo {author} {\bibfnamefont
  {W.}~\bibnamefont {Lu}},\ and\ \bibinfo {author} {\bibfnamefont
  {M.}~\bibnamefont {Tian}},\ }\bibfield  {title} {\bibinfo {title} {Evidences
  of topological surface states in the nodal-line semimetal sntas2
  nanoflakes},\ }\href {https://doi.org/10.1021/acsnano.2c11932} {\bibfield
  {journal} {\bibinfo  {journal} {ACS Nano}\ }\textbf {\bibinfo {volume}
  {17}},\ \bibinfo {pages} {4913} (\bibinfo {year} {2023})},\ \bibinfo {note}
  {pMID: 36802534},\ \Eprint
  {https://arxiv.org/abs/https://doi.org/10.1021/acsnano.2c11932}
  {https://doi.org/10.1021/acsnano.2c11932} \BibitemShut {NoStop}%
\bibitem [{\citenamefont {Abdulla}\ \emph
  {et~al.}(2023{\natexlab{a}})\citenamefont {Abdulla}, \citenamefont {Murthy},\
  and\ \citenamefont {Das}}]{Abdulla_SNS_2023}%
  \BibitemOpen
  \bibfield  {author} {\bibinfo {author} {\bibfnamefont {F.}~\bibnamefont
  {Abdulla}}, \bibinfo {author} {\bibfnamefont {G.}~\bibnamefont {Murthy}},\
  and\ \bibinfo {author} {\bibfnamefont {A.}~\bibnamefont {Das}},\ }\href
  {https://arxiv.org/abs/2401.02966} {\bibinfo {title} {Stable nodal line
  semimetals in the chiral classes in three dimensions}} (\bibinfo {year}
  {2023}{\natexlab{a}}),\ \Eprint {https://arxiv.org/abs/2401.02966}
  {arXiv:2401.02966 [cond-mat.mes-hall]} \BibitemShut {NoStop}%
\bibitem [{\citenamefont {Abdulla}\ \emph
  {et~al.}(2023{\natexlab{b}})\citenamefont {Abdulla}, \citenamefont {Murthy},\
  and\ \citenamefont {Das}}]{Abdulla_TNL_2023}%
  \BibitemOpen
  \bibfield  {author} {\bibinfo {author} {\bibfnamefont {F.}~\bibnamefont
  {Abdulla}}, \bibinfo {author} {\bibfnamefont {G.}~\bibnamefont {Murthy}},\
  and\ \bibinfo {author} {\bibfnamefont {A.}~\bibnamefont {Das}},\ }\href
  {https://arxiv.org/abs/2311.18667} {\bibinfo {title} {Topological nodal line
  semimetals with chiral symmetry}} (\bibinfo {year} {2023}{\natexlab{b}}),\
  \Eprint {https://arxiv.org/abs/2311.18667} {arXiv:2311.18667
  [cond-mat.mes-hall]} \BibitemShut {NoStop}%
\bibitem [{\citenamefont {Abdulla}\ \emph {et~al.}(2025)\citenamefont
  {Abdulla}, \citenamefont {Murthy},\ and\ \citenamefont
  {Das}}]{Abdulla_ISP_2025}%
  \BibitemOpen
  \bibfield  {author} {\bibinfo {author} {\bibfnamefont {F.}~\bibnamefont
  {Abdulla}}, \bibinfo {author} {\bibfnamefont {G.}~\bibnamefont {Murthy}},\
  and\ \bibinfo {author} {\bibfnamefont {A.}~\bibnamefont {Das}},\ }\bibfield
  {title} {\bibinfo {title} {Internal symmetry protected stable topological
  semimetals in three dimensions},\ }\href {https://doi.org/10.1103/q4gd-b4vb}
  {\bibfield  {journal} {\bibinfo  {journal} {Phys. Rev. B}\ }\textbf {\bibinfo
  {volume} {112}},\ \bibinfo {pages} {205111} (\bibinfo {year}
  {2025})}\BibitemShut {NoStop}%
\bibitem [{\citenamefont {Zhao}\ and\ \citenamefont
  {Wang}(2013)}]{Zhao_TCS_2013}%
  \BibitemOpen
  \bibfield  {author} {\bibinfo {author} {\bibfnamefont {Y.~X.}\ \bibnamefont
  {Zhao}}\ and\ \bibinfo {author} {\bibfnamefont {Z.~D.}\ \bibnamefont
  {Wang}},\ }\bibfield  {title} {\bibinfo {title} {Topological classification
  and stability of fermi surfaces},\ }\href
  {https://doi.org/10.1103/PhysRevLett.110.240404} {\bibfield  {journal}
  {\bibinfo  {journal} {Phys. Rev. Lett.}\ }\textbf {\bibinfo {volume} {110}},\
  \bibinfo {pages} {240404} (\bibinfo {year} {2013})}\BibitemShut {NoStop}%
\bibitem [{\citenamefont {Matsuura}\ \emph {et~al.}(2013)\citenamefont
  {Matsuura}, \citenamefont {Chang}, \citenamefont {Schnyder},\ and\
  \citenamefont {Ryu}}]{Matsuura_PBS_2013}%
  \BibitemOpen
  \bibfield  {author} {\bibinfo {author} {\bibfnamefont {S.}~\bibnamefont
  {Matsuura}}, \bibinfo {author} {\bibfnamefont {P.-Y.}\ \bibnamefont {Chang}},
  \bibinfo {author} {\bibfnamefont {A.~P.}\ \bibnamefont {Schnyder}},\ and\
  \bibinfo {author} {\bibfnamefont {S.}~\bibnamefont {Ryu}},\ }\bibfield
  {title} {\bibinfo {title} {Protected boundary states in gapless topological
  phases},\ }\href {https://doi.org/10.1088/1367-2630/15/6/065001} {\bibfield
  {journal} {\bibinfo  {journal} {New Journal of Physics}\ }\textbf {\bibinfo
  {volume} {15}},\ \bibinfo {pages} {065001} (\bibinfo {year}
  {2013})}\BibitemShut {NoStop}%
\bibitem [{\citenamefont {Rudi}\ \emph {et~al.}(2024)\citenamefont {Rudi},
  \citenamefont {De~Martino}, \citenamefont {Moors}, \citenamefont {Giuliano},\
  and\ \citenamefont {Buccheri}}]{Buccheri_Interface_2024}%
  \BibitemOpen
  \bibfield  {author} {\bibinfo {author} {\bibfnamefont {M.}~\bibnamefont
  {Rudi}}, \bibinfo {author} {\bibfnamefont {A.}~\bibnamefont {De~Martino}},
  \bibinfo {author} {\bibfnamefont {K.}~\bibnamefont {Moors}}, \bibinfo
  {author} {\bibfnamefont {D.}~\bibnamefont {Giuliano}},\ and\ \bibinfo
  {author} {\bibfnamefont {F.}~\bibnamefont {Buccheri}},\ }\bibfield  {title}
  {\bibinfo {title} {Interfaces of nodal-line semimetals: Drum states,
  transport, and refraction},\ }\href
  {https://doi.org/10.1103/PhysRevB.109.195144} {\bibfield  {journal} {\bibinfo
   {journal} {Phys. Rev. B}\ }\textbf {\bibinfo {volume} {109}},\ \bibinfo
  {pages} {195144} (\bibinfo {year} {2024})}\BibitemShut {NoStop}%
\bibitem [{\citenamefont {Okamoto}\ \emph {et~al.}(2014)\citenamefont
  {Okamoto}, \citenamefont {Takane},\ and\ \citenamefont
  {Imura}}]{Okamoto_Onedimensional_2014}%
  \BibitemOpen
  \bibfield  {author} {\bibinfo {author} {\bibfnamefont {M.}~\bibnamefont
  {Okamoto}}, \bibinfo {author} {\bibfnamefont {Y.}~\bibnamefont {Takane}},\
  and\ \bibinfo {author} {\bibfnamefont {K.-I.}\ \bibnamefont {Imura}},\
  }\bibfield  {title} {\bibinfo {title} {One-dimensional topological insulator:
  A model for studying finite-size effects in topological insulator thin
  films},\ }\href {https://doi.org/10.1103/PhysRevB.89.125425} {\bibfield
  {journal} {\bibinfo  {journal} {Phys. Rev. B}\ }\textbf {\bibinfo {volume}
  {89}},\ \bibinfo {pages} {125425} (\bibinfo {year} {2014})}\BibitemShut
  {NoStop}%
\bibitem [{\citenamefont {Wu}\ \emph {et~al.}(2018)\citenamefont {Wu},
  \citenamefont {Liu}, \citenamefont {Li}, \citenamefont {Zhong}, \citenamefont
  {Yu}, \citenamefont {Sheng}, \citenamefont {Zhao},\ and\ \citenamefont
  {Yang}}]{Wu_2018}%
  \BibitemOpen
  \bibfield  {author} {\bibinfo {author} {\bibfnamefont {W.}~\bibnamefont
  {Wu}}, \bibinfo {author} {\bibfnamefont {Y.}~\bibnamefont {Liu}}, \bibinfo
  {author} {\bibfnamefont {S.}~\bibnamefont {Li}}, \bibinfo {author}
  {\bibfnamefont {C.}~\bibnamefont {Zhong}}, \bibinfo {author} {\bibfnamefont
  {Z.-M.}\ \bibnamefont {Yu}}, \bibinfo {author} {\bibfnamefont {X.-L.}\
  \bibnamefont {Sheng}}, \bibinfo {author} {\bibfnamefont {Y.~X.}\ \bibnamefont
  {Zhao}},\ and\ \bibinfo {author} {\bibfnamefont {S.~A.}\ \bibnamefont
  {Yang}},\ }\bibfield  {title} {\bibinfo {title} {Nodal surface semimetals:
  Theory and material realization},\ }\href
  {https://doi.org/10.1103/PhysRevB.97.115125} {\bibfield  {journal} {\bibinfo
  {journal} {Phys. Rev. B}\ }\textbf {\bibinfo {volume} {97}},\ \bibinfo
  {pages} {115125} (\bibinfo {year} {2018})}\BibitemShut {NoStop}%
\bibitem [{\citenamefont {Nguyen}\ \emph {et~al.}(2021)\citenamefont {Nguyen},
  \citenamefont {Kobayashi}, \citenamefont {Wichmann},\ and\ \citenamefont
  {Nomura}}]{Kentaro_Quantum_2021}%
  \BibitemOpen
  \bibfield  {author} {\bibinfo {author} {\bibfnamefont {D.-H.-M.}\
  \bibnamefont {Nguyen}}, \bibinfo {author} {\bibfnamefont {K.}~\bibnamefont
  {Kobayashi}}, \bibinfo {author} {\bibfnamefont {J.-E.~R.}\ \bibnamefont
  {Wichmann}},\ and\ \bibinfo {author} {\bibfnamefont {K.}~\bibnamefont
  {Nomura}},\ }\bibfield  {title} {\bibinfo {title} {Quantum hall effect
  induced by chiral landau levels in topological semimetal films},\ }\href
  {https://doi.org/10.1103/PhysRevB.104.045302} {\bibfield  {journal} {\bibinfo
   {journal} {Phys. Rev. B}\ }\textbf {\bibinfo {volume} {104}},\ \bibinfo
  {pages} {045302} (\bibinfo {year} {2021})}\BibitemShut {NoStop}%
\bibitem [{\citenamefont {Isobe}\ and\ \citenamefont
  {Nagaosa}(2016)}]{Isobe_Nagaosa_2016}%
  \BibitemOpen
  \bibfield  {author} {\bibinfo {author} {\bibfnamefont {H.}~\bibnamefont
  {Isobe}}\ and\ \bibinfo {author} {\bibfnamefont {N.}~\bibnamefont
  {Nagaosa}},\ }\bibfield  {title} {\bibinfo {title} {Coulomb interaction
  effect in weyl fermions with tilted energy dispersion in two dimensions},\
  }\href {https://doi.org/10.1103/PhysRevLett.116.116803} {\bibfield  {journal}
  {\bibinfo  {journal} {Phys. Rev. Lett.}\ }\textbf {\bibinfo {volume} {116}},\
  \bibinfo {pages} {116803} (\bibinfo {year} {2016})}\BibitemShut {NoStop}%
\bibitem [{\citenamefont {Meng}\ \emph {et~al.}(2021)\citenamefont {Meng},
  \citenamefont {Zhang}, \citenamefont {Liu}, \citenamefont {Wang},
  \citenamefont {Dai},\ and\ \citenamefont {Liu}}]{Meng_Zhang_2021}%
  \BibitemOpen
  \bibfield  {author} {\bibinfo {author} {\bibfnamefont {W.}~\bibnamefont
  {Meng}}, \bibinfo {author} {\bibfnamefont {X.}~\bibnamefont {Zhang}},
  \bibinfo {author} {\bibfnamefont {Y.}~\bibnamefont {Liu}}, \bibinfo {author}
  {\bibfnamefont {L.}~\bibnamefont {Wang}}, \bibinfo {author} {\bibfnamefont
  {X.}~\bibnamefont {Dai}},\ and\ \bibinfo {author} {\bibfnamefont
  {G.}~\bibnamefont {Liu}},\ }\bibfield  {title} {\bibinfo {title}
  {Two-dimensional weyl semimetal with coexisting fully spin-polarized type-i
  and type-ii weyl points},\ }\href
  {https://doi.org/https://doi.org/10.1016/j.apsusc.2020.148318} {\bibfield
  {journal} {\bibinfo  {journal} {Applied Surface Science}\ }\textbf {\bibinfo
  {volume} {540}},\ \bibinfo {pages} {148318} (\bibinfo {year}
  {2021})}\BibitemShut {NoStop}%
\bibitem [{\citenamefont {Shi}\ \emph {et~al.}(2021)\citenamefont {Shi},
  \citenamefont {Li}, \citenamefont {Cui}, \citenamefont {Song},\ and\
  \citenamefont {Liu}}]{Shi_Liu_2021}%
  \BibitemOpen
  \bibfield  {author} {\bibinfo {author} {\bibfnamefont {Y.}~\bibnamefont
  {Shi}}, \bibinfo {author} {\bibfnamefont {L.}~\bibnamefont {Li}}, \bibinfo
  {author} {\bibfnamefont {X.}~\bibnamefont {Cui}}, \bibinfo {author}
  {\bibfnamefont {T.}~\bibnamefont {Song}},\ and\ \bibinfo {author}
  {\bibfnamefont {Z.}~\bibnamefont {Liu}},\ }\bibfield  {title} {\bibinfo
  {title} {Mnnbr monolayer: A high-temperature ferromagnetic half-metal with
  type-ii weyl fermions},\ }\href
  {https://doi.org/https://doi.org/10.1002/pssr.202100115} {\bibfield
  {journal} {\bibinfo  {journal} {physica status solidi (RRL) – Rapid
  Research Letters}\ }\textbf {\bibinfo {volume} {15}},\ \bibinfo {pages}
  {2100115} (\bibinfo {year} {2021})},\ \Eprint
  {https://arxiv.org/abs/https://onlinelibrary.wiley.com/doi/pdf/10.1002/pssr.202100115}
  {https://onlinelibrary.wiley.com/doi/pdf/10.1002/pssr.202100115} \BibitemShut
  {NoStop}%
\bibitem [{\citenamefont {Li}\ \emph {et~al.}(2021)\citenamefont {Li},
  \citenamefont {Xie}, \citenamefont {Ding}, \citenamefont {Ji}, \citenamefont
  {Li}, \citenamefont {Zhang}, \citenamefont {Li},\ and\ \citenamefont
  {Wang}}]{Li_Wang_2021}%
  \BibitemOpen
  \bibfield  {author} {\bibinfo {author} {\bibfnamefont {G.-G.}\ \bibnamefont
  {Li}}, \bibinfo {author} {\bibfnamefont {R.-R.}\ \bibnamefont {Xie}},
  \bibinfo {author} {\bibfnamefont {L.-J.}\ \bibnamefont {Ding}}, \bibinfo
  {author} {\bibfnamefont {W.-X.}\ \bibnamefont {Ji}}, \bibinfo {author}
  {\bibfnamefont {S.-S.}\ \bibnamefont {Li}}, \bibinfo {author} {\bibfnamefont
  {C.-W.}\ \bibnamefont {Zhang}}, \bibinfo {author} {\bibfnamefont
  {P.}~\bibnamefont {Li}},\ and\ \bibinfo {author} {\bibfnamefont {P.-J.}\
  \bibnamefont {Wang}},\ }\bibfield  {title} {\bibinfo {title} {Two-dimensional
  weyl semi-half-metallic nics3 with a band structure controllable by the
  direction of magnetization},\ }\href {https://doi.org/10.1039/D1CP00812A}
  {\bibfield  {journal} {\bibinfo  {journal} {Phys. Chem. Chem. Phys.}\
  }\textbf {\bibinfo {volume} {23}},\ \bibinfo {pages} {12068} (\bibinfo {year}
  {2021})}\BibitemShut {NoStop}%
\bibitem [{\citenamefont {Wei}\ \emph {et~al.}(2022)\citenamefont {Wei},
  \citenamefont {Yang}, \citenamefont {Shen},\ and\ \citenamefont
  {Tao}}]{Wei_Tao_2022}%
  \BibitemOpen
  \bibfield  {author} {\bibinfo {author} {\bibfnamefont {X.-P.}\ \bibnamefont
  {Wei}}, \bibinfo {author} {\bibfnamefont {N.}~\bibnamefont {Yang}}, \bibinfo
  {author} {\bibfnamefont {J.}~\bibnamefont {Shen}},\ and\ \bibinfo {author}
  {\bibfnamefont {X.}~\bibnamefont {Tao}},\ }\bibfield  {title} {\bibinfo
  {title} {Ferromagnetic weyl semimetals and quantum anomalous hall effect in
  2d half-metallic mn2nt2},\ }\href
  {https://doi.org/https://doi.org/10.1016/j.physe.2022.115164} {\bibfield
  {journal} {\bibinfo  {journal} {Physica E: Low-dimensional Systems and
  Nanostructures}\ }\textbf {\bibinfo {volume} {140}},\ \bibinfo {pages}
  {115164} (\bibinfo {year} {2022})}\BibitemShut {NoStop}%
\bibitem [{\citenamefont {Jia}\ \emph {et~al.}(2020)\citenamefont {Jia},
  \citenamefont {Meng}, \citenamefont {Zhang}, \citenamefont {Liu},
  \citenamefont {Dai}, \citenamefont {Zhang},\ and\ \citenamefont
  {Liu}}]{Jia_Liu_2020}%
  \BibitemOpen
  \bibfield  {author} {\bibinfo {author} {\bibfnamefont {T.}~\bibnamefont
  {Jia}}, \bibinfo {author} {\bibfnamefont {W.}~\bibnamefont {Meng}}, \bibinfo
  {author} {\bibfnamefont {H.}~\bibnamefont {Zhang}}, \bibinfo {author}
  {\bibfnamefont {C.}~\bibnamefont {Liu}}, \bibinfo {author} {\bibfnamefont
  {X.}~\bibnamefont {Dai}}, \bibinfo {author} {\bibfnamefont {X.}~\bibnamefont
  {Zhang}},\ and\ \bibinfo {author} {\bibfnamefont {G.}~\bibnamefont {Liu}},\
  }\bibfield  {title} {\bibinfo {title} {Weyl fermions in vi3 monolayer},\
  }\bibfield  {journal} {\bibinfo  {journal} {Frontiers in Chemistry}\ }\textbf
  {\bibinfo {volume} {8}},\ \href {https://doi.org/10.3389/fchem.2020.00722}
  {10.3389/fchem.2020.00722} (\bibinfo {year} {2020})\BibitemShut {NoStop}%
\bibitem [{\citenamefont {You}\ \emph {et~al.}(2019)\citenamefont {You},
  \citenamefont {Chen}, \citenamefont {Zhang}, \citenamefont {Sheng},
  \citenamefont {Yang},\ and\ \citenamefont {Su}}]{You_Su_2019}%
  \BibitemOpen
  \bibfield  {author} {\bibinfo {author} {\bibfnamefont {J.-Y.}\ \bibnamefont
  {You}}, \bibinfo {author} {\bibfnamefont {C.}~\bibnamefont {Chen}}, \bibinfo
  {author} {\bibfnamefont {Z.}~\bibnamefont {Zhang}}, \bibinfo {author}
  {\bibfnamefont {X.-L.}\ \bibnamefont {Sheng}}, \bibinfo {author}
  {\bibfnamefont {S.~A.}\ \bibnamefont {Yang}},\ and\ \bibinfo {author}
  {\bibfnamefont {G.}~\bibnamefont {Su}},\ }\bibfield  {title} {\bibinfo
  {title} {Two-dimensional weyl half-semimetal and tunable quantum anomalous
  hall effect},\ }\href {https://doi.org/10.1103/PhysRevB.100.064408}
  {\bibfield  {journal} {\bibinfo  {journal} {Phys. Rev. B}\ }\textbf {\bibinfo
  {volume} {100}},\ \bibinfo {pages} {064408} (\bibinfo {year}
  {2019})}\BibitemShut {NoStop}%
\bibitem [{\citenamefont {Lopes}\ \emph {et~al.}(2024)\citenamefont {Lopes},
  \citenamefont {Baierle}, \citenamefont {Miwa},\ and\ \citenamefont
  {Schmidt}}]{Lopes_2024}%
  \BibitemOpen
  \bibfield  {author} {\bibinfo {author} {\bibfnamefont {E.~V.~C.}\
  \bibnamefont {Lopes}}, \bibinfo {author} {\bibfnamefont {R.~J.}\ \bibnamefont
  {Baierle}}, \bibinfo {author} {\bibfnamefont {R.~H.}\ \bibnamefont {Miwa}},\
  and\ \bibinfo {author} {\bibfnamefont {T.~M.}\ \bibnamefont {Schmidt}},\
  }\bibfield  {title} {\bibinfo {title} {Noncentrosymmetric two-dimensional
  weyl semimetals in porous si/ge structures},\ }\href
  {https://doi.org/10.1088/1361-648X/ad1e09} {\bibfield  {journal} {\bibinfo
  {journal} {Journal of Physics: Condensed Matter}\ }\textbf {\bibinfo {volume}
  {36}},\ \bibinfo {pages} {185701} (\bibinfo {year} {2024})}\BibitemShut
  {NoStop}%
\bibitem [{\citenamefont {Guo}\ \emph {et~al.}(2023{\natexlab{b}})\citenamefont
  {Guo}, \citenamefont {Miao}, \citenamefont {Huang}, \citenamefont {Lygo},
  \citenamefont {Dai},\ and\ \citenamefont {Stemmer}}]{Stemmer_2023}%
  \BibitemOpen
  \bibfield  {author} {\bibinfo {author} {\bibfnamefont {B.}~\bibnamefont
  {Guo}}, \bibinfo {author} {\bibfnamefont {W.}~\bibnamefont {Miao}}, \bibinfo
  {author} {\bibfnamefont {V.}~\bibnamefont {Huang}}, \bibinfo {author}
  {\bibfnamefont {A.~C.}\ \bibnamefont {Lygo}}, \bibinfo {author}
  {\bibfnamefont {X.}~\bibnamefont {Dai}},\ and\ \bibinfo {author}
  {\bibfnamefont {S.}~\bibnamefont {Stemmer}},\ }\bibfield  {title} {\bibinfo
  {title} {Zeeman field-induced two-dimensional weyl semimetal phase in cadmium
  arsenide},\ }\href {https://doi.org/10.1103/PhysRevLett.131.046601}
  {\bibfield  {journal} {\bibinfo  {journal} {Phys. Rev. Lett.}\ }\textbf
  {\bibinfo {volume} {131}},\ \bibinfo {pages} {046601} (\bibinfo {year}
  {2023}{\natexlab{b}})}\BibitemShut {NoStop}%
\bibitem [{\citenamefont {Abdulla}(2024)}]{Abdulla_Protected_2024}%
  \BibitemOpen
  \bibfield  {author} {\bibinfo {author} {\bibfnamefont {F.}~\bibnamefont
  {Abdulla}},\ }\href {https://arxiv.org/abs/2401.04656} {\bibinfo {title}
  {Protected weyl semimetals within 2d chiral classes}} (\bibinfo {year}
  {2024}),\ \Eprint {https://arxiv.org/abs/2401.04656} {arXiv:2401.04656
  [cond-mat.mes-hall]} \BibitemShut {NoStop}%
\bibitem [{\citenamefont {Nielsen}\ and\ \citenamefont
  {Ninomiya}(1983)}]{Nielsen_Ninomiya_1983}%
  \BibitemOpen
  \bibfield  {author} {\bibinfo {author} {\bibfnamefont {H.}~\bibnamefont
  {Nielsen}}\ and\ \bibinfo {author} {\bibfnamefont {M.}~\bibnamefont
  {Ninomiya}},\ }\bibfield  {title} {\bibinfo {title} {The adler-bell-jackiw
  anomaly and weyl fermions in a crystal},\ }\href
  {https://doi.org/https://doi.org/10.1016/0370-2693(83)91529-0} {\bibfield
  {journal} {\bibinfo  {journal} {Physics Letters B}\ }\textbf {\bibinfo
  {volume} {130}},\ \bibinfo {pages} {389} (\bibinfo {year}
  {1983})}\BibitemShut {NoStop}%
\bibitem [{\citenamefont {Muechler}\ \emph {et~al.}(2020)\citenamefont
  {Muechler}, \citenamefont {Topp}, \citenamefont {Queiroz}, \citenamefont
  {Krivenkov}, \citenamefont {Varykhalov}, \citenamefont {Cano}, \citenamefont
  {Ast},\ and\ \citenamefont {Schoop}}]{Lukas_Modular_2020}%
  \BibitemOpen
  \bibfield  {author} {\bibinfo {author} {\bibfnamefont {L.}~\bibnamefont
  {Muechler}}, \bibinfo {author} {\bibfnamefont {A.}~\bibnamefont {Topp}},
  \bibinfo {author} {\bibfnamefont {R.}~\bibnamefont {Queiroz}}, \bibinfo
  {author} {\bibfnamefont {M.}~\bibnamefont {Krivenkov}}, \bibinfo {author}
  {\bibfnamefont {A.}~\bibnamefont {Varykhalov}}, \bibinfo {author}
  {\bibfnamefont {J.}~\bibnamefont {Cano}}, \bibinfo {author} {\bibfnamefont
  {C.~R.}\ \bibnamefont {Ast}},\ and\ \bibinfo {author} {\bibfnamefont {L.~M.}\
  \bibnamefont {Schoop}},\ }\bibfield  {title} {\bibinfo {title} {Modular
  arithmetic with nodal lines: Drumhead surface states in zrsite},\ }\href
  {https://doi.org/10.1103/PhysRevX.10.011026} {\bibfield  {journal} {\bibinfo
  {journal} {Phys. Rev. X}\ }\textbf {\bibinfo {volume} {10}},\ \bibinfo
  {pages} {011026} (\bibinfo {year} {2020})}\BibitemShut {NoStop}%
\end{thebibliography}%

\end{document}